\documentclass[Manuscript,12pt]{elsarticle}
\usepackage{graphicx,stfloats}
\usepackage{color}

\biboptions{sort&compress}

\journal{Combustion \& Flame}

\usepackage[normalem]{ulem}
\usepackage[english]{babel}
\usepackage{caption}
\usepackage{fontenc}
\usepackage{color, colortbl}
\usepackage{multirow}
\usepackage[pdftex]{hyperref}
\usepackage{subfigure}
\usepackage{amsmath}
\usepackage{subeqnarray}
\usepackage[all]{xy}
\usepackage{ctable}
\usepackage{stfloats}
 \usepackage{dcolumn}
\usepackage{array}
\usepackage{graphicx}
\hypersetup{pdftitle={}}
\usepackage{natbib}
\usepackage{footnote}
\usepackage{amssymb}
\usepackage{calrsfs}

\usepackage{hyperref}

\hypersetup{
	    pdfproducer={},  
	    pdfcreator={},
}

\journal{Combustion and flame}

\begin{document}

\begin{frontmatter}
\title{Experimental Analysis of Oscillatory Premixed Flames in a Hele-Shaw Cell Propagating Towards a Closed End}
\author[uc3m]{Fernando Veiga-L\'opez}
\author[uc3m]{Daniel Mart\'inez-Ruiz}
\author[uc3m]{Eduardo Fern\'andez-Tarrazo}
\author[uc3m]{Mario S\'anchez-Sanz \corref{cor1}}
\address[uc3m]{Departamento de Ingenier\'{\i}a T\'ermica y de Fluidos, Universidad Carlos 
	III de Madrid, 28911, Legan\'es, Madrid, Spain}
\cortext[cor1]{mssanz@ing.uc3m.es}


\begin{abstract}
An experimental study of methane, propane and dimethyl ether (DME) premixed flames propagating in a quasi-two-dimensional Hele-Shaw cell placed horizontally is presented in this paper. The flames are ignited at the open end of the combustion chamber and propagate towards the closed end. Our experiments revealed two distinct propagation regimes depending on the equivalence ratio of the mixture as a consequence of the coupling between the heat-release rate and the acoustic waves. The primary acoustic instability induces a small-amplitude, of around 8 mm, oscillatory motion across the chamber that is observed for lean propane, lean DME, and rich methane flames. Eventually, a secondary acoustic instability emerges for sufficiently rich (lean) propane and DME (methane) flames, inducing large-amplitude oscillations in the direction of propagation of the flame. The amplitude of these oscillations can be as large as 30 mm and drastically changes the outline of the flame. The front then forms pulsating finger-shaped structures that characterize the flame propagation under the secondary acoustic instability. \\

The experimental setup allows the recording of the flame propagation from two different points of view. The top view is used to obtain accurate quantitative information about the flame propagation, while the lateral view offered a novel three dimensional perspective of the flame that gives relevant information on the transition between the two oscillatory regimes.\\ 

The influence of the geometry of the Hele-Shaw cell and of the equivalence ratio on the transition between the two acoustic-instability regimes is analyzed. In particular, we find that the transition to the secondary instability occurs for values of the equivalence ratio $\phi$ above (below) a critical value $\phi_c$ for propane and DME (methane) flames. In all the tested fuels, the transition to the secondary instability emerges for values of the Markstein number ${\cal M}$ below a critical value ${\cal M}_c$. The critical Markstein number varies  with the gap size $h$ formed by the two horizontal plates that bound the Hele-Shaw cell. As $h$ is reduced, the critical Markstein number is shifted towards larger values. 

\end{abstract}

\end{frontmatter}

\ifdefined \wordcount
\clearpage
\fi

\section{Introduction}
The work by Searby \cite{searby1992acoustic} described experimentally the development of primary and secondary acoustic instabilities in propane flames traveling downwards in a tube with the ignition end open to the atmosphere. 
In a different experiment, Aldredge and Killingsworth \cite{aldredge2004experimental} tested a premixed methane flame propagating downwards in a Taylor-Couette burner. In their experiments with methane, they found that rich flames were more stable than lean flames, a  similar  behavior to that of the propane flames reported by Searby. In turn, Ya\~nez et al. \cite{yanez2015flame} repeated the experiment using hydrogen mixtures  to find oscillating flames only for very lean mixtures. \\

Connelly and Kyritsis \cite{connelly2014} and Yang et al. \cite{yang2015oscillating} carried out experiments with propane flames propagating along narrow tubes open at both ends. As in the experiments by Searby, the flames propagate with large and small-amplitude oscillations depending on the stoichiometry of the mixture. 
Almarcha et al. \cite{almarcha2015experimental}, Gross and Pan \cite{Gross2014} and Shariff et al. \cite{sharif1999premixed} studied experimentally flames of propane and hydrogen propagating in a Hele-Shaw cell open at the ignition end. None of these papers reported any oscillatory regime, neither primary nor secondary, in their experiments.\\ 

Since the work on pyro-acoustic interaction presented in \cite{searby1991parametric}, it seems clear that the secondary acoustic instability is caused by pre-existing finite-amplitude acoustic oscillations and, therefore, a significant amount of work has been dedicated to explain their origin \cite{clavin1990one,clanet1999primary}. Two mechanisms were considered to explain the generation of the primary acoustic instability in tubes: the direct sensitivity of the reaction rate to acoustic pressure and the variation of the flame-front area induced by acoustic acceleration.  A detailed account of the progress on flame thermoacoustic instabilities in tubes can be found in the book by Clavin and Searby \cite{clavin2016combustion}. 
Recently, Yoon et al. \cite{yoon2016onset} found a correlation between the product $\beta M$, being $\beta$ and $M$ the Zel'dovich and the Mach number respectively, and the average acoustic pressure. Their work suggests that the sensitivity of the reaction rate to acoustic pressure dominates the process. Moreover, the interaction between the coupling constant $\beta M$  with the Lewis number was examined in \cite{yoon2017effects} and \cite{yoon2018experimental}.\\
 	
The existence of two different oscillatory regimes, attributed to flame-acoustics resonance, in a flame propagating towards the closed end in a narrow channel has been reported numerically by Petchenko et al. \cite{petchenko2006violent}. According to their results, the acoustic oscillations produce an effective acceleration field at the flame front leading to a strong Rayleigh-Taylor instability which intensely wrinkles the flame-front. Later, in a paper by the same author \cite{petchenko2007flame}, the effect of the parameter $h/\delta_T$ was included, being $h$ the channel height and $\delta_T=D_T/S_L$ the flame thickness, with $D_T$ the thermal diffusivity of the gas and $S_L$ the planar flame burning velocity. According to their results, the oscillations of the flame become stronger in wider domains, inducing flame folding in sufficiently wide tubes. On the other hand, small-amplitude flame oscillations were obtained in their calculation even in very narrow tubes $h/\delta_T =10$. However, Kurdyumov and Matalon \cite{Kurdyumov2016} found a non-oscillatory propagation speed for a flame advancing towards the closed end when solving the same problem numerically in the limit of very narrow channels $h/\delta_T \ll 1$ and including the effect of gas compressibility in their formulation. Fern\'andez-Galisteo et al. \cite{Galisteo2018} used the low-Mach number approximation $M \ll 1$ to compute numerically a flame propagating in a Hele-Shaw cell in the limit $h/\delta_T \ll 1$. They found instabilities that wrinkled the flame increasing the surface area and the propagation speed but, since they neglected the compressibility of the gas, they could not reproduce the acoustic variations affecting the flame.\\

Most of the experimental and theoretical studies found in the literature focused their attention on flames propagating in tubes. We present in this paper an experimental study in a quasi-two-dimensional Hele-Shaw cell to contribute to the understanding of the transition between the primary and secondary instabilities for different fuels. Moreover, the role played by the equivalence ratio and the channel height-to-flame thickness ratio ($h/\delta_T$) in the interaction between the acoustic waves and the flame-front is investigated by changing the geometry of the combustion chamber and the mixture composition respectively. \\

\section{Experimental setup and procedure} 
\label{sec:experimental_procedure}
\begin{figure}[!t]
	\centering
	\includegraphics[width=0.75\textwidth]{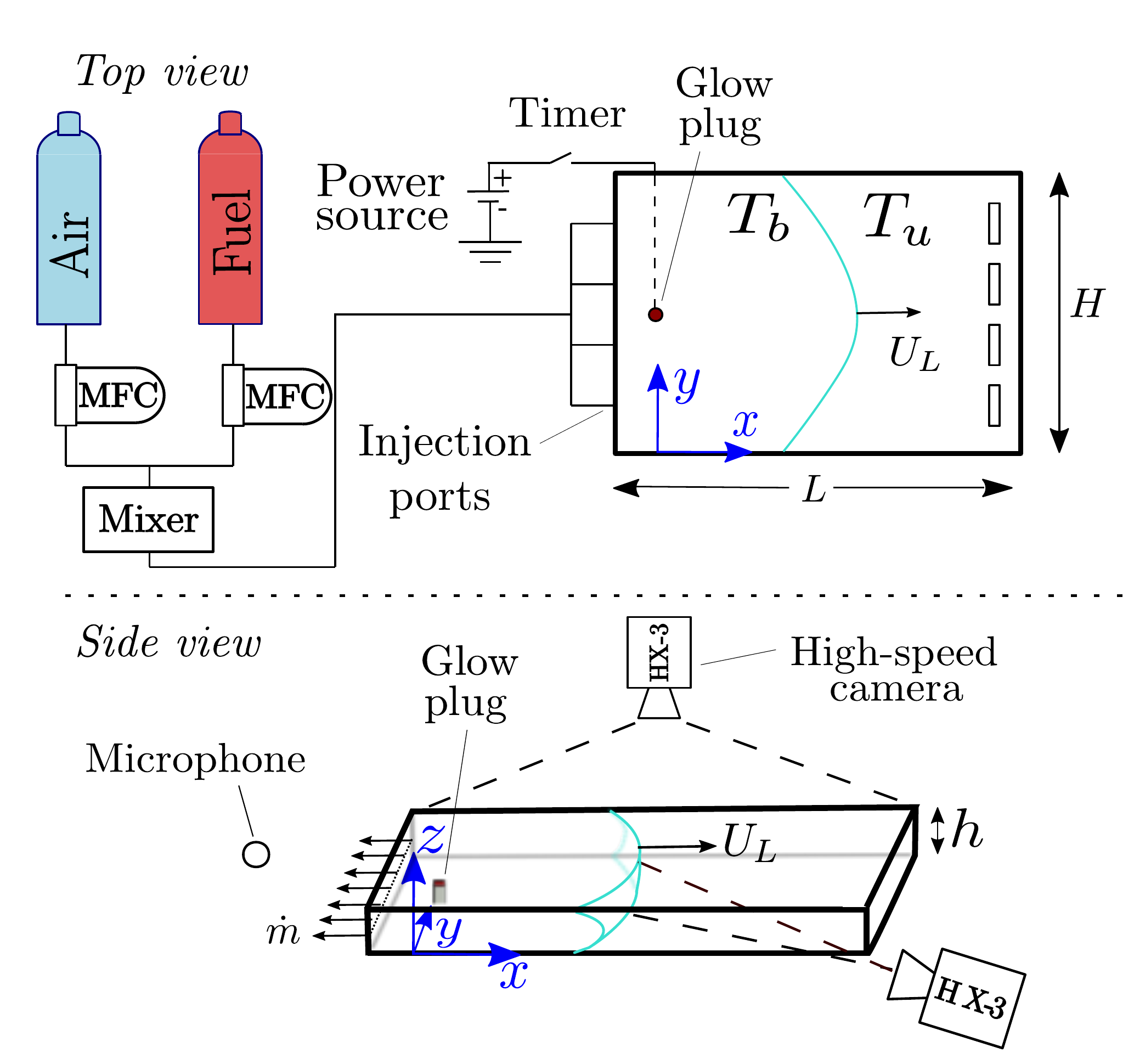}
	\caption{Schematic representation of the experimental apparatus. In the figure we depict the top and side view of the Hele-Shaw cell indicating its dimensions $L \times H \times h$, the definition of the coordinates system, the location of the glow plug and the different positions of the high-speed camera to record the flame from two different points of view. In the figure, the black arrow lines located at the ignition end indicate the flow $\dot{m}$  of high-temperature gas off the Hele-Shaw cell.}
	\label{fig:sketch}
\end{figure}

The experimental setup is sketched in Fig.~\ref{fig:sketch}. The combustion chamber is formed by two flat plates disposed horizontally and separated by a PMMA hollow frame, enclosing a maximum volume of $L \times H \times h =900 \times 500 \times 10$ mm$^3$. Nevertheless, the observation length of the chamber from the glow plug to the opposite end is 800 mm.
The top cover is a 19~mm-thick tempered-glass plate while the lower one is a rigid aluminum table insulated with a 1~mm vinyl layer. The gap between the plates can be varied from 10 mm to 1 mm by staking 3 mm-thick PVC laminae inside the hollow frame. The chamber is filled with a fuel-air mixture which is prepared before injection using two mass flow controllers (Sierra SmartTrak 100 for fuel and Omega FMA5418A, 0-5 slm for air) to regulate the equivalence ratio $\phi$.\\
 
The experimental procedure starts by injecting the mixture into the chamber at the ignition end, via four regularly-spaced injection ports. During the charge of the reactants, the ignition end of the vessel is kept sealed while the opposite end is opened for venting. Upon the complete charge, when the stopcock is closed, both ends are shut to allow the gases to come to rest. Then, the ignition-injection end is reopened and the mixture is ignited using a glow plug (BOSCH Duraspeed) powered with an amount of electrical energy that is held constant for all the experiments. The whole section $H \times h$ at the ignition end is available to freely vent the high-temperature products off the chamber. No valve nor gas extraction device was used, avoiding any possible disruption of the exhaust gases outflow.\\

The luminous emission of the flame is recorded with a high-speed camera (MEMRECAM HX-3) shooting at 4000 fps, if not specified otherwise. The experimental setup allows shooting videos from two points of view to capture both the top and side views of the flames, as shown in Fig. \ref{fig:sketch}. The top view has been used to obtain accurate quantitative data from the recording (such as oscillation frequencies, burned volume fraction, flame velocity, etc.). The side view offers a novel three dimensional perspective of the reactive front that reveals important features of the flame propagation.  Simultaneously, the acoustic signal is recorded using a microphone located at the open ignition end. Image, audio and post-processing analyses are performed using an in-house Python code.\\
\begin{table}
	\centering
	\caption{Properties of the fuel-air mixtures calculated at room temperature $T_u=291$ K with $R_g=286.99 $ m$^2$/(s$^2$ K), $E=125400$ J/mol \cite{westbrook1981simplified} for C$_3$H$_8$ (propane), $E=132129$ J/mol \cite{fernandez2006simple} for CH$_4$ (methane) and $E=250800$ J/mol for  CH$_3$OCH$_3$ (DME), calculated by fitting the experimental values  of the planar flame speed $S_L$ measured by \cite{de2011laminar}, to the flame speed calculated using an Arrhenius expression for the reaction rate in the form $\Omega= A [F] [O]\exp \left\{-E/R T\right\}$, with $A= 1\times 10^{31} $ in cm$^3$/mol s, $R=8.31$ J/mol K and $[F]$ and $[O]$ representing the concentration in mol/cm$^3$ of fuel and oxidizer. The adiabatic flame temperature $T_b$ and the planar flame speed $S_L$ are calculated using the San Diego mechanism. For DME flames, the values of ${\cal M}$ for $\phi<0.7$ are extrapolated. The Mach number is defined as $M=S_L/c$, with the sound speed measured at room temperature. The thermal flame thickness is given by $\delta_T=D_T/S_L$ with $D_T=2\times 10^{-5}$ m$^2$/s. The small-amplitude oscillations are observed in flames with the equivalence ratios highlighted in gray.}        
\definecolor{airforceblue}{rgb}{0.74, 0.83, 0.9}
\definecolor{airforceblue}{gray}{0.7}
\label{table:flame_parameters}
	{\footnotesize
	\begin{tabular}{c|cccccccccc|}
		\toprule
         \rowcolor{white}
		 & $\phi$ & $T_b$   & $S_L$ & $\delta_T$ & $Le_F$ & $Le_O$ & $Le_{eff}$ & $\beta$ & $\beta  M$ & $\cal{M}$ \\ 
		 &  & [K]   & [cm/s] & [mm] &  &  &  &  & $\times 10^{3}$   &  \\ \midrule
		\multirow{8}{*}{{\rotatebox[origin=c]{90}{C$_3$H$_8$}}}           
        & \cellcolor{airforceblue}0.70   & \cellcolor{airforceblue}1884.80 & \cellcolor{airforceblue}21.95 &\cellcolor{airforceblue} 0.09       & \cellcolor{airforceblue}1.85   & \cellcolor{airforceblue}1.06   & \cellcolor{airforceblue}1.66       & \cellcolor{airforceblue}7.10     & \cellcolor{airforceblue}4.50       & \cellcolor{airforceblue}3.16      \\
		&\cellcolor{airforceblue}0.80   & \cellcolor{airforceblue}2048.10 & \cellcolor{airforceblue}28.91 & \cellcolor{airforceblue}0.07       & \cellcolor{airforceblue}1.84   & \cellcolor{airforceblue}1.06   & \cellcolor{airforceblue}1.61       & \cellcolor{airforceblue}6.63     & \cellcolor{airforceblue}5.54       & \cellcolor{airforceblue}2.98      \\
        & \cellcolor{airforceblue}0.87   & \cellcolor{airforceblue}2143.76 & \cellcolor{airforceblue}33.17 & \cellcolor{airforceblue}0.06       & \cellcolor{airforceblue}1.83   & \cellcolor{airforceblue}1.06   & \cellcolor{airforceblue}1.55       & \cellcolor{airforceblue}6.08     &\cellcolor{airforceblue} 6.00       & \cellcolor{airforceblue}2.92      \\
		& 0.90   & 2184.90 & 35.00 & 0.06       & 1.83   & 1.05   & 1.53       & 6.28     & 6.35       & 2.79      \\
		& 1.00   & 2271.50 & 38.91 & 0.05       & 1.82   & 1.05   & 1.44       & 6.08     & 6.84       & 2.35      \\
		& 1.10   & 2268.10 & 40.50 & 0.05       & 1.81   & 1.05   & 1.34       & 6.09     & 7.12       & 1.96      \\
		& 1.20   & 2200.60 & 38.62 & 0.05       & 1.80   & 1.04   & 1.27       & 6.24     & 6.97       & 1.77      \\		 \midrule
		\multirow{8}{*}{{\rotatebox[origin=c]{90}{CH$_4$}}}         
        & 0.70   & 1833.71 & 15.41 & 0.13       & 0.98   & 1.10   & 1.01       & 7.26     & 3.23       & 0.75      \\
		& 0.80   & 1992.31 & 24.12 & 0.08       & 0.98   & 1.10   & 1.02       & 6.78     & 4.73       & 0.77      \\
		& 0.90   & 2130.22 & 32.50 & 0.06       & 0.98   & 1.10   & 1.03       & 6.42     & 6.03       & 0.81      \\
        & \cellcolor{airforceblue}0.95   & \cellcolor{airforceblue}2176.19 & \cellcolor{airforceblue}34.79 & \cellcolor{airforceblue}0.06       & \cellcolor{airforceblue}0.98   & \cellcolor{airforceblue}1.09   & \cellcolor{airforceblue}1.03       & \cellcolor{airforceblue}6.33     & \cellcolor{airforceblue}6.03       & \cellcolor{airforceblue}0.81      \\
		& \cellcolor{airforceblue}1.00   &\cellcolor{airforceblue} 2222.16 & \cellcolor{airforceblue}37.08 & \cellcolor{airforceblue}0.05       &\cellcolor{airforceblue} 0.98   & \cellcolor{airforceblue}1.08   & \cellcolor{airforceblue}1.03       & \cellcolor{airforceblue}6.19     & \cellcolor{airforceblue}6.64       & \cellcolor{airforceblue}0.88      \\
		& \cellcolor{airforceblue}1.10   & \cellcolor{airforceblue} 2206.32 & \cellcolor{airforceblue} 35.01 & \cellcolor{airforceblue} 0.06       & \cellcolor{airforceblue} 0.98   & \cellcolor{airforceblue} 1.08   & \cellcolor{airforceblue} 1.04       & \cellcolor{airforceblue} 6.23     & \cellcolor{airforceblue} 6.30       & \cellcolor{airforceblue} 0.96      \\
		& \cellcolor{airforceblue} 1.20   & \cellcolor{airforceblue} 2132.02 & \cellcolor{airforceblue} 32.18 & \cellcolor{airforceblue} 0.06       & \cellcolor{airforceblue} 0.98   & \cellcolor{airforceblue} 1.09   & \cellcolor{airforceblue} 1.06       & \cellcolor{airforceblue} 6.41     & \cellcolor{airforceblue} 5.96       & \cellcolor{airforceblue} 1.02      \\\midrule
		\multirow{13}{*}{{\rotatebox[origin=c]{90}{ CH$_3$OCH$_3$}}}  
        & \cellcolor{airforceblue}0.50   & \cellcolor{airforceblue}1563.20 & \cellcolor{airforceblue}7.09  & \cellcolor{airforceblue}0.28       & \cellcolor{airforceblue}1.82   &\cellcolor{airforceblue} 1.06   & \cellcolor{airforceblue}1.71       &\cellcolor{airforceblue} 10.41    & \cellcolor{airforceblue}2.13       &   \cellcolor{airforceblue} 4.09   \\
        & \cellcolor{airforceblue}0.55	 & \cellcolor{airforceblue}1669.40 & \cellcolor{airforceblue}11.74 & \cellcolor{airforceblue}0.19   	& \cellcolor{airforceblue}1.81	 & \cellcolor{airforceblue}1.06   &	\cellcolor{airforceblue}1.69	   & \cellcolor{airforceblue}9.91     &	\cellcolor{airforceblue}3.31       &   \cellcolor{airforceblue}  3.90 \\
		& 0.60   & 1756.30 & 15.55 & 0.13       & 1.80   & 1.06   & 1.67       & 9.51     & 4.27       & 3.72      \\
		& 0.70   & 1932.60 & 24.79 & 0.08       & 1.79   & 1.05   & 1.63       & 8.80     & 6.31       & 3.35      \\
		& 0.80   & 2088.00 & 33.06 & 0.06       & 1.78   & 1.04   & 1.58       & 8.26     & 7.89       & 2.98      \\
		& 0.90   & 2212.60 & 39.62 & 0.05       & 1.76   & 1.04   & 1.50       & 7.86     & 9.01       & 2.39      \\
		& 1.00   & 2289.80 & 44.04 & 0.05       & 1.75   & 1.03   & 1.39       & 7.64     & 9.72       & 1.86      \\
		& 1.10   & 2296.20 & 45.97 & 0.04       & 1.74   & 1.03   & 1.28       & 7.62     & 10.13      & 1.75      \\
		& 1.20   & 2246.00 & 45.13 & 0.04       & 1.73   & 1.02   & 1.22       & 7.77     & 10.13      & 1.61     \\
		  \bottomrule
	\end{tabular} }
\end{table}\\

	A summary of the properties of the flames tested in the experiments is included in Table \ref{table:flame_parameters}. This table encompasses the planar burning velocity $S_L$, the flame thickness $\delta_T= D_T/S_L$ and the adiabatic flame temperature at equilibrium for propane, methane  and DME, which have been calculated using COSILAB. In the same table, we include the coupling parameter $\beta M$, identified in \cite{yoon2016onset,yoon2017effects} as a key quantity controlling the formation of the primary acoustic instability. Here $\beta=E(T_b-T_u)/R T_b^2$ is the Zel'dovich number, calculated using the activation energy $E$, the adiabatic flame temperature $T_b$ and the room temperature $T_u$, and $M=S_L/c$, is the Mach number, with $c$ the speed of sound of the fresh mixture. \\

With the idea of contributing to the understanding of the transition from the primary to the secondary instability described in \cite{searby1992acoustic}, we also added in Table~\ref{table:flame_parameters} the effective Lewis number $Le_{eff}$  and the Markstein number $\mathcal{M}$, proposed by \cite{aldredge2004experimental} as an important parameter in explaining the effect of acoustic fluctuations on local flame stretch. The Markstein number ${\cal M}=\mathcal{L}/\delta_T $ is defined as the ratio between the Markstein length $\mathcal{L}$ and the flame thickness $\delta_T$, and quantifies the effects of curvature and strain $\mathbb{K}$ on the propagation velocity as
\begin{equation*}
S_f=S_L - \mathcal{L} \mathbb{K} 
\end{equation*}
The values of the Markstein number are taken from \cite{bechtold2001dependence} for methane and propane and from \cite{de2011laminar} for DME.\\ 

The effective Lewis number $L_{eff}$, introduced by  \cite{joulin1981linear} and \cite{garcia1984soret}, which controls the amplification of the hydrodynamic instability due to diffusion effects \cite{joulin1981linear}, is calculated as  $Le_{eff}=(Le_O+(1-\tilde{\phi})Le_F)/(2-\tilde{\phi})$ for lean flames and  $Le_{eff}=(Le_F+(1+\tilde{\phi})Le_O))/(2+\tilde{\phi})$ for rich flames, with $\tilde{\phi}=\beta(\phi-1)$, where the subscripts $O$ and $F$ refer to oxidizer and fuel respectively \cite{bechtold2001dependence}. 

\section{Experimental results }

\subsection{Propane (C$_3$H$_8$) flames}
\begin{figure*}[!ht]
	\centering
	\begin{tabular}[t]{cc}
		\includegraphics[width=0.5\textwidth]{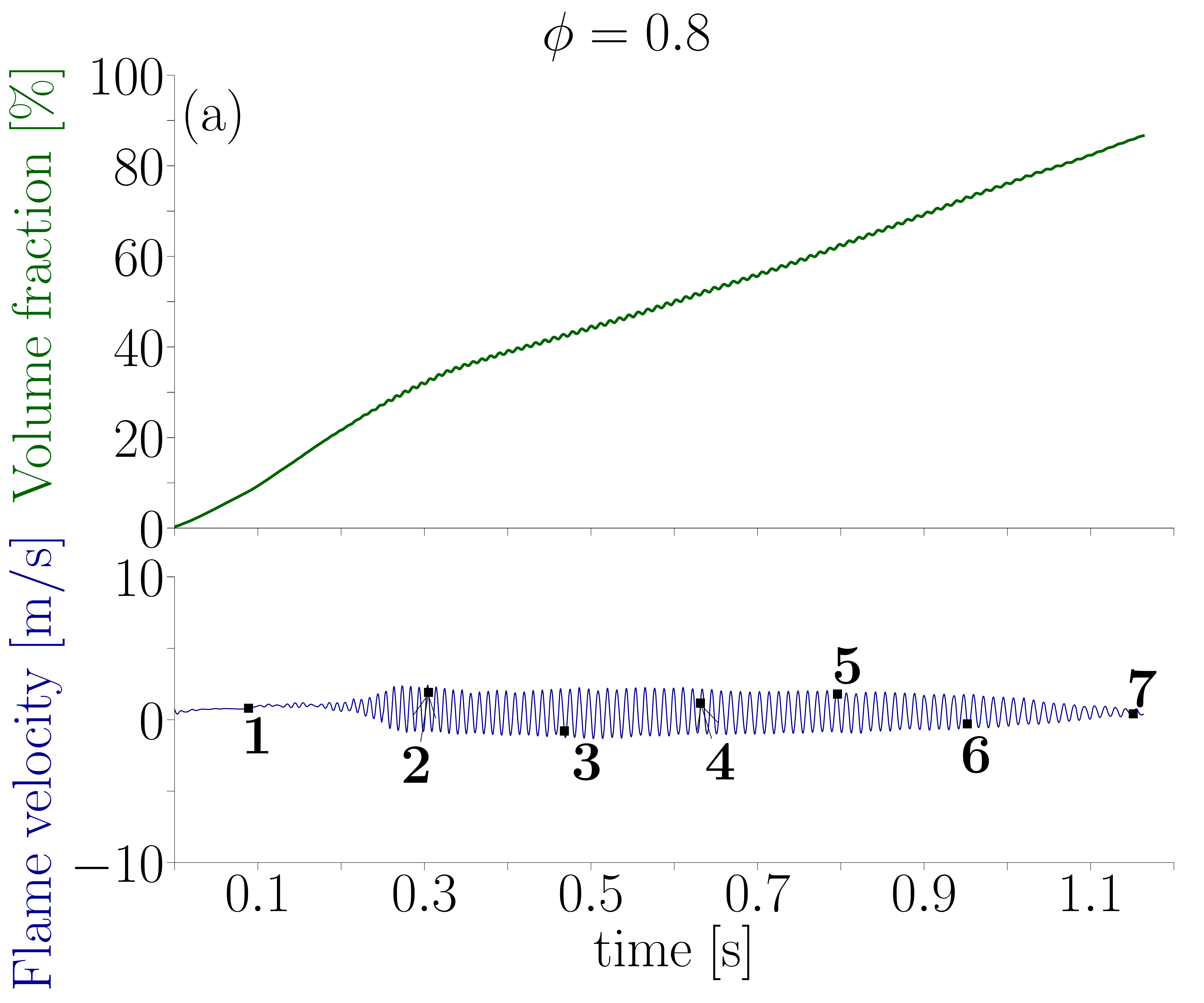} &
		\includegraphics[width=0.5\textwidth]{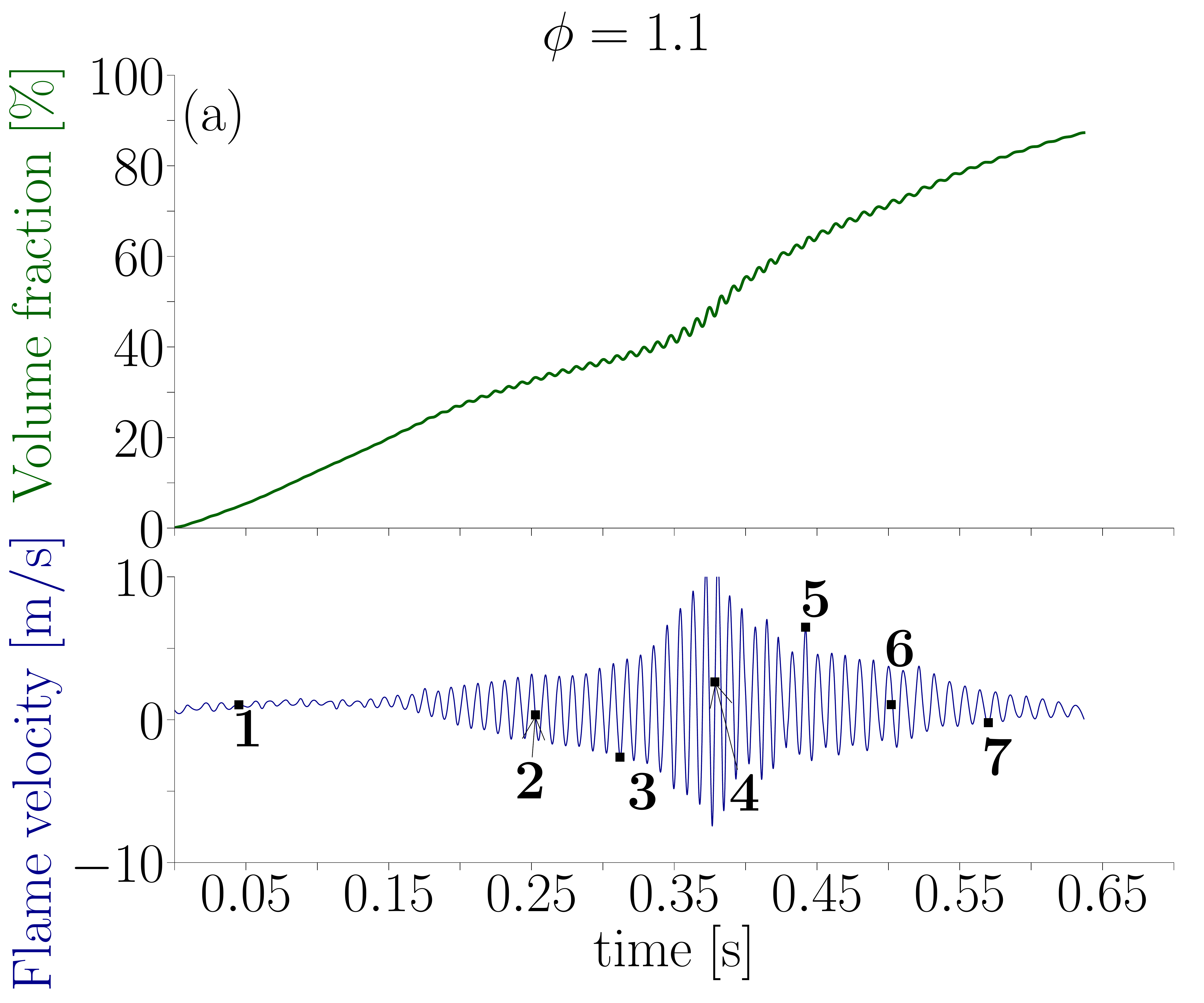}	\\		
		\includegraphics[width=0.5\textwidth]{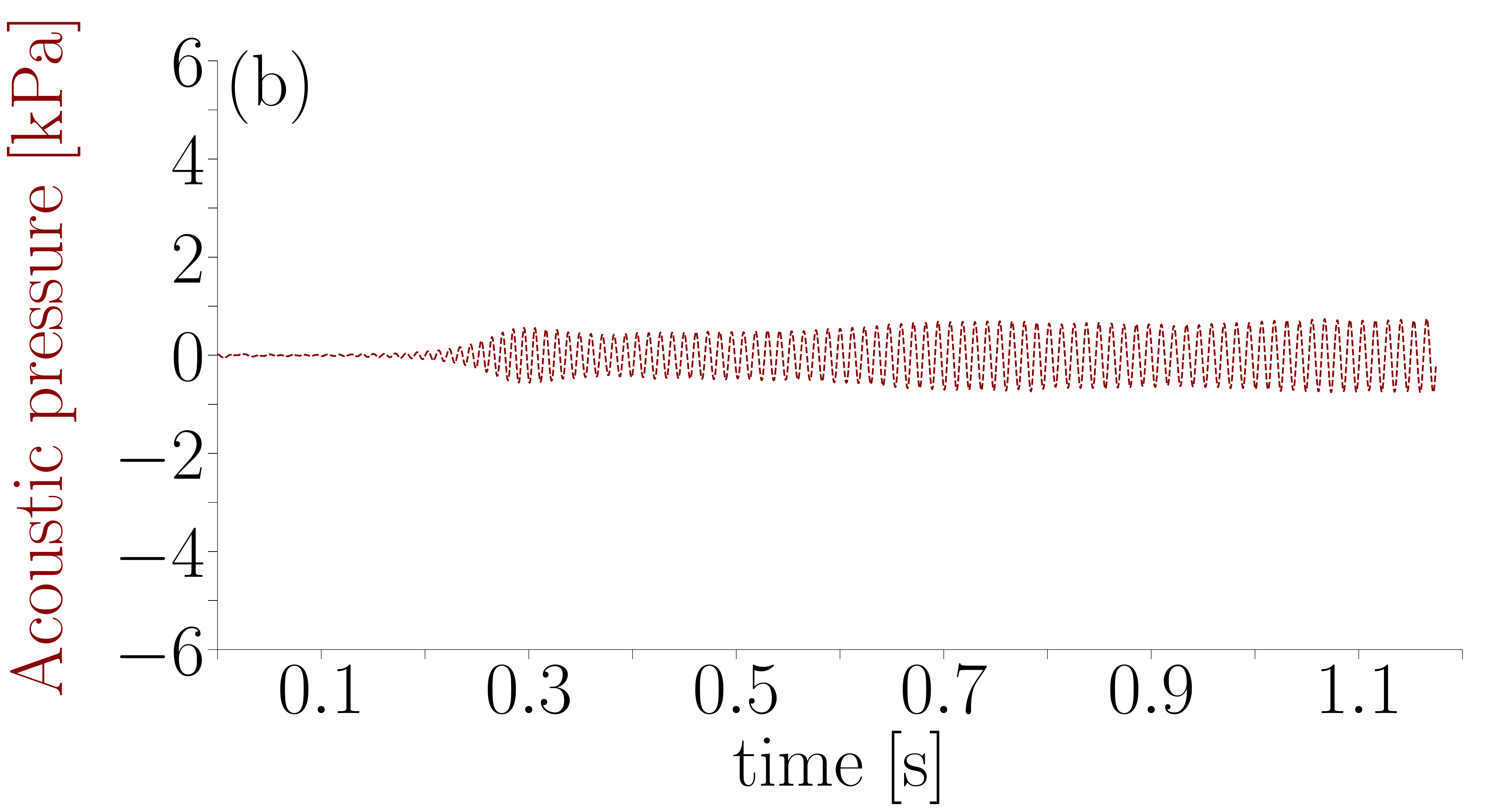}&				
		\includegraphics[width=0.5\textwidth]{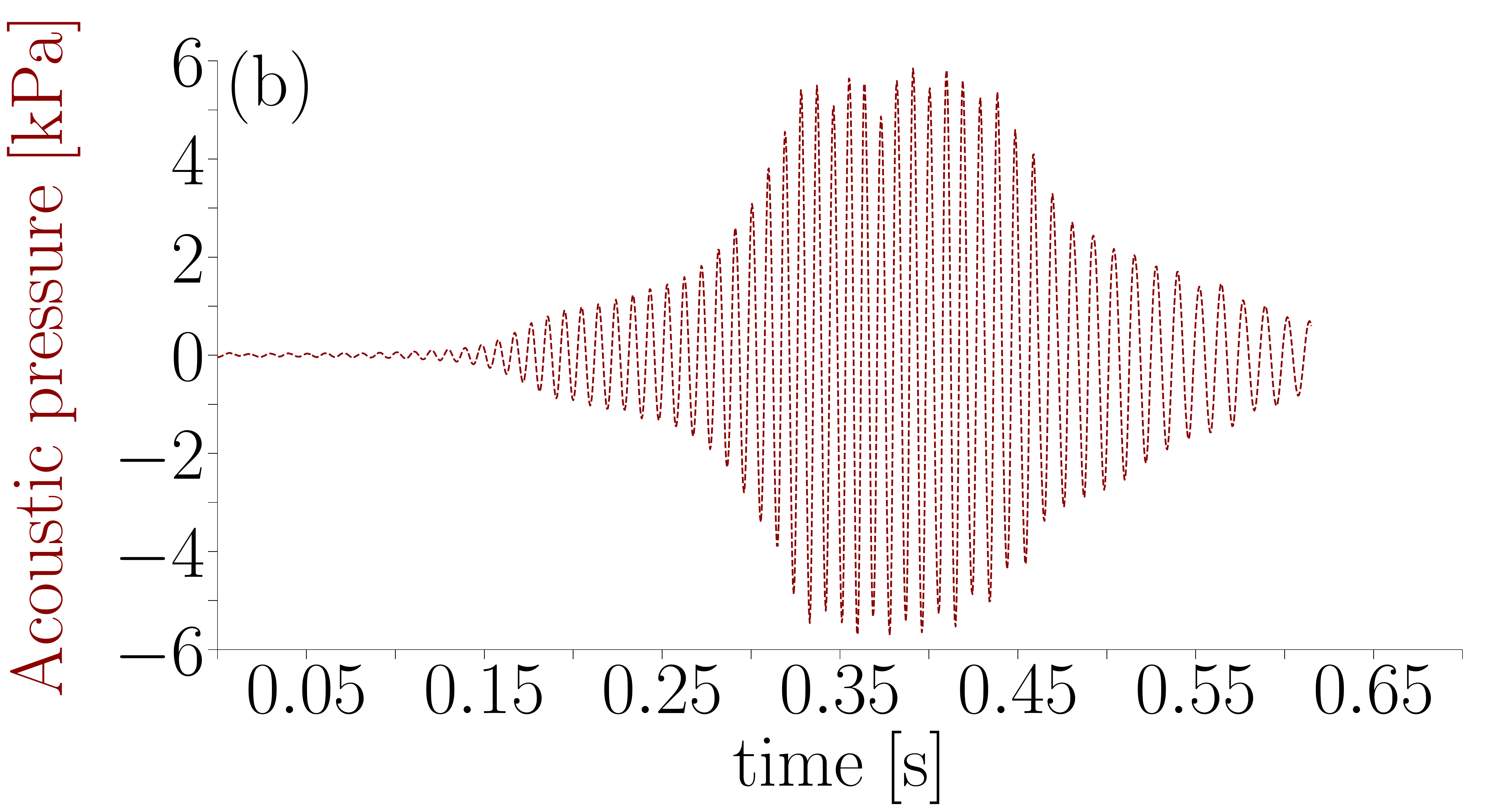}
	\end{tabular}	\caption{\textbf{(a)} Time evolution of the relative burned volume fraction $V_b/V$ (upper thick-solid lines) swept by a propane-air flame and the flame velocity calculated from the burned volume  $U_L=(Hh)^{-1} dV_b/dt$  (lower thin-solid lines). The error in the determination of the equivalence ratio is $\pm 0.05$ and the maximum uncertainty of the burned volume measurements is $\pm 2.75\%$. \textbf{(b)} Acoustic pressure as a function of time.}
	\label{fig:composed}
\end{figure*}

The aforementioned methodology is first applied here to the widely studied propane-air mixtures. The evolution with time of both the burned volume and the flame velocity are plotted versus time in Fig. \ref{fig:composed}. The  burned volume fraction of burned gases is defined here as the ratio between the volume of the chamber occupied by the high-temperature gas $V_b$, obtained from the images, and the total chamber volume $V = H \times L \times h$. Furthermore, the flame velocity of the flame is calculated assuming a flat flame with the same burned volume as observed in the experiments $U_L=(Hh)^{-1} dV_b/dt$.\\

The data plotted in this figure illustrate the two different oscillatory behaviors measured in our experiments. The lean flame, left panels of Fig. \ref{fig:composed}, propagates along the Hele-Shaw cell with a propagation velocity that oscillates around a positive mean value. The amplitude of the oscillation is small, of around 8 mm, and remains stable until the flame reaches the end of the chamber.\\

Contrarily, the amplitude of the flame oscillation is as large as 30 mm for rich mixtures, inducing changes of around 2.5\% in the burned volume fraction, with peak velocities close to 10 m/s, as obtained via image analysis in Fig.~\ref{fig:composed}a (right). Similarly to the experiments by Searby \cite{searby1992acoustic}, the average propagation velocity increases during this phase, proved by the increment of the slope at approximately halfway of the chamber in Fig. \ref{fig:composed}a (right). The transition between oscillatory regimes takes place suddenly at a critical equivalence ratio measured to be $\phi_c=0.87 \pm 0.05$. \\

The oscillations of the flame can be compared to the acoustic pressure registered by the microphone, plotted in Fig.~\ref{fig:composed}b. After an initial period, during which the microphone records the ignition event, the device measures the sound generated by the flame. This figure shows a sudden increase in the amplitude of the acoustic pressure that becomes an order of magnitude larger for $\phi =1.1$  ($\Delta p \sim 6 $ kPa) than for $\phi=0.8$ ($\Delta p \sim 0.5$ kPa). The match between the signal from the microphone and the flame oscillations suggests a coupling between the sound waves propagating within the chamber and the behavior of the flame that will be further examined below in section \ref{sec:freq}.\\

\begin{figure}[!ht]
	\centering
	\includegraphics[width=0.85\textwidth,angle=-0]{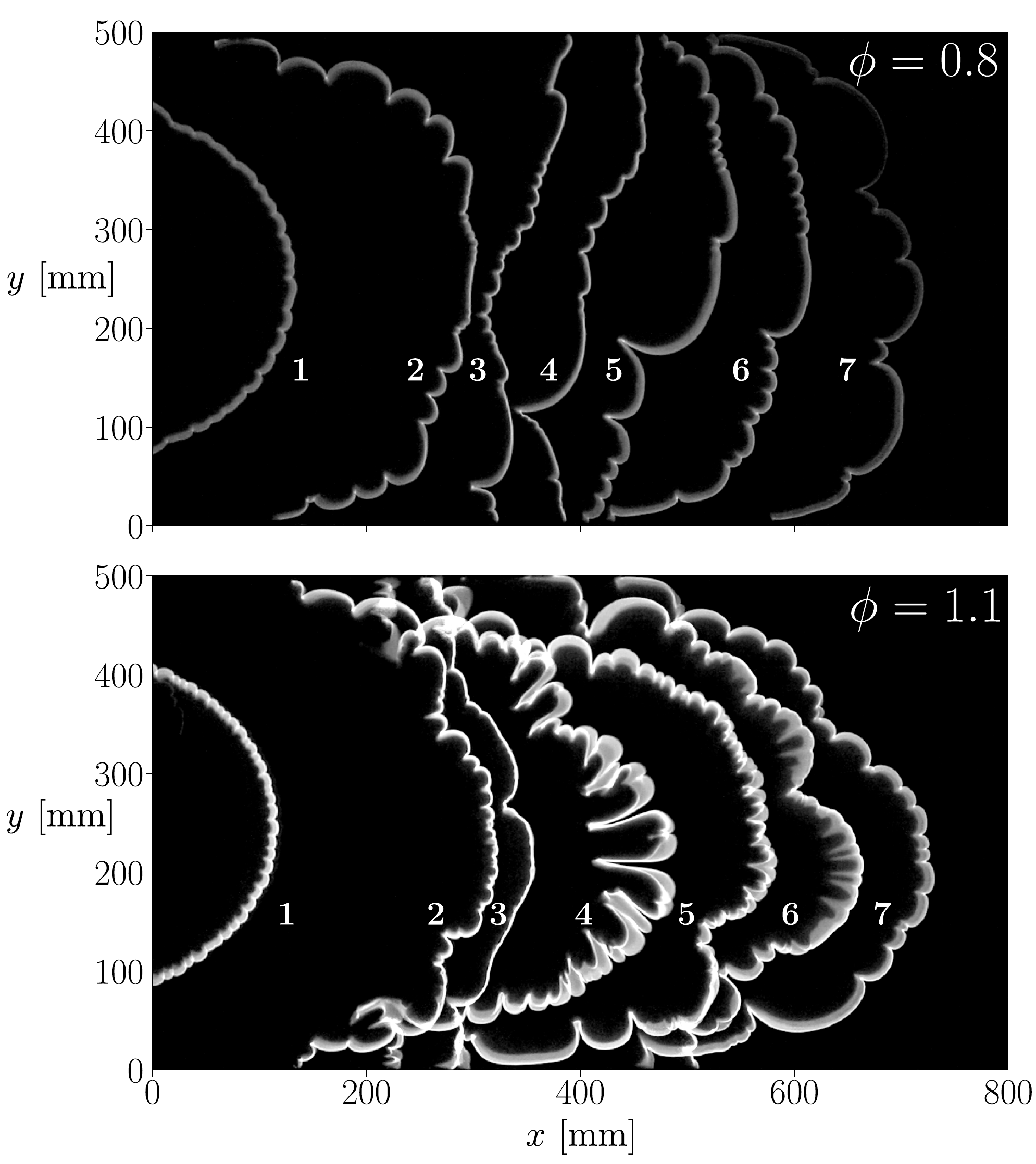} 
	\caption{Luminous emissions of the flame recorded by the high-speed camera at the times indicated by the numbers 1 to 7 included in Fig.~\ref{fig:composed}.}
	\label{fig:top}
\end{figure}

\begin{figure}[!t]
	\centering
	\includegraphics[width=0.85\textwidth,angle=-0]{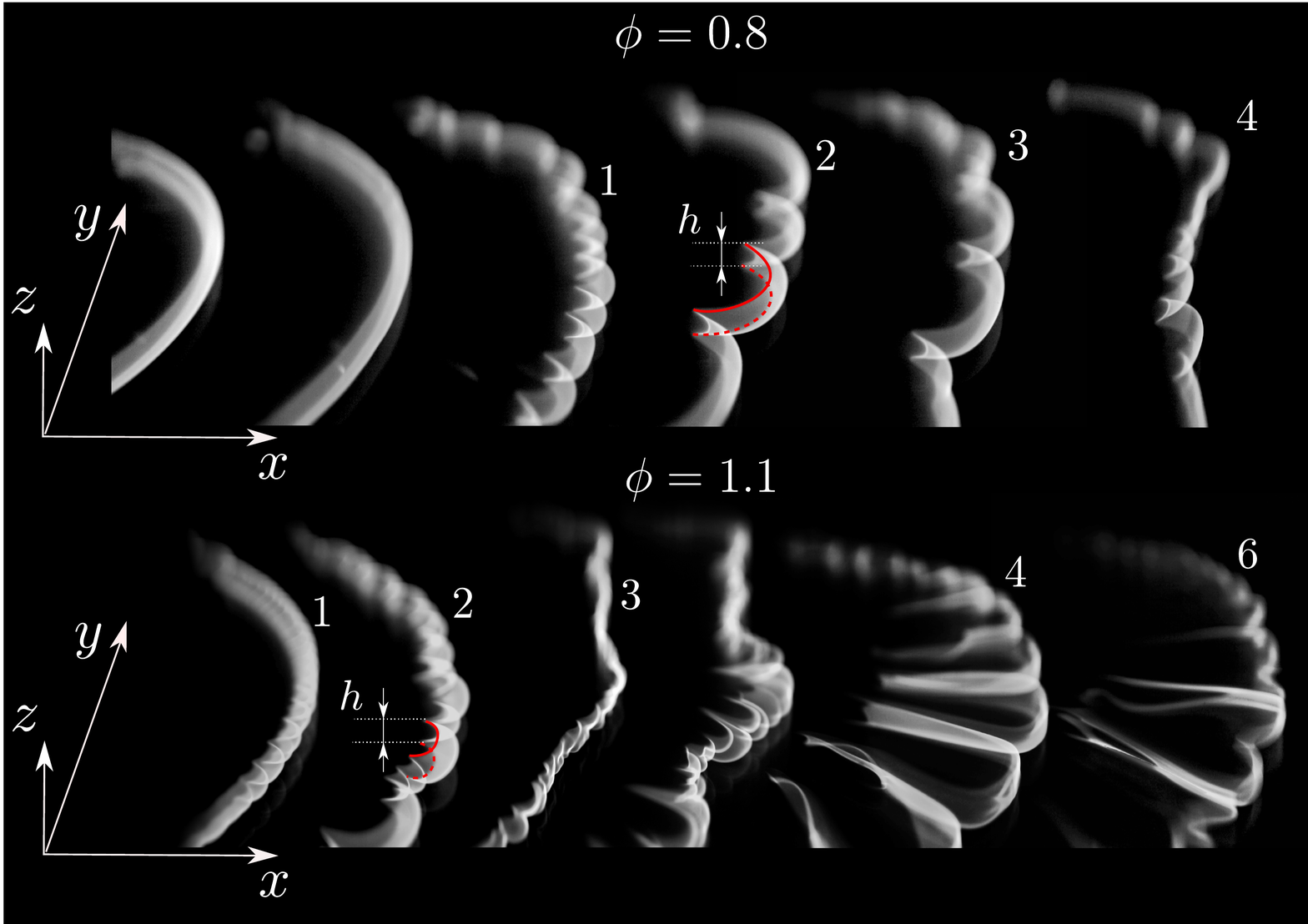}
	\caption{Side view, with the camera placed laterally as indicated in the sketch of Fig. \ref{fig:sketch}, of the propane flame with $\phi=0.8$ and $\phi=1.1$ and a cell thickness of $h=10$ mm. The images displayed in this figure and those shown in Fig. \ref{fig:top} are taken in  different runs of the experiment. The numbers in the image indicate that the flame is approximately at the same stage within the Helle-Shaw cell than the flame with the same number in Fig. \ref{fig:top}. The solid and dashed red lines indicate where the flame touches the upper and lower horizontal plates, respectively.}
	\label{fig:lateral}
\end{figure}

\begin{figure}[!t]
	\centering
	\includegraphics[width=0.7\textwidth,angle=-0]{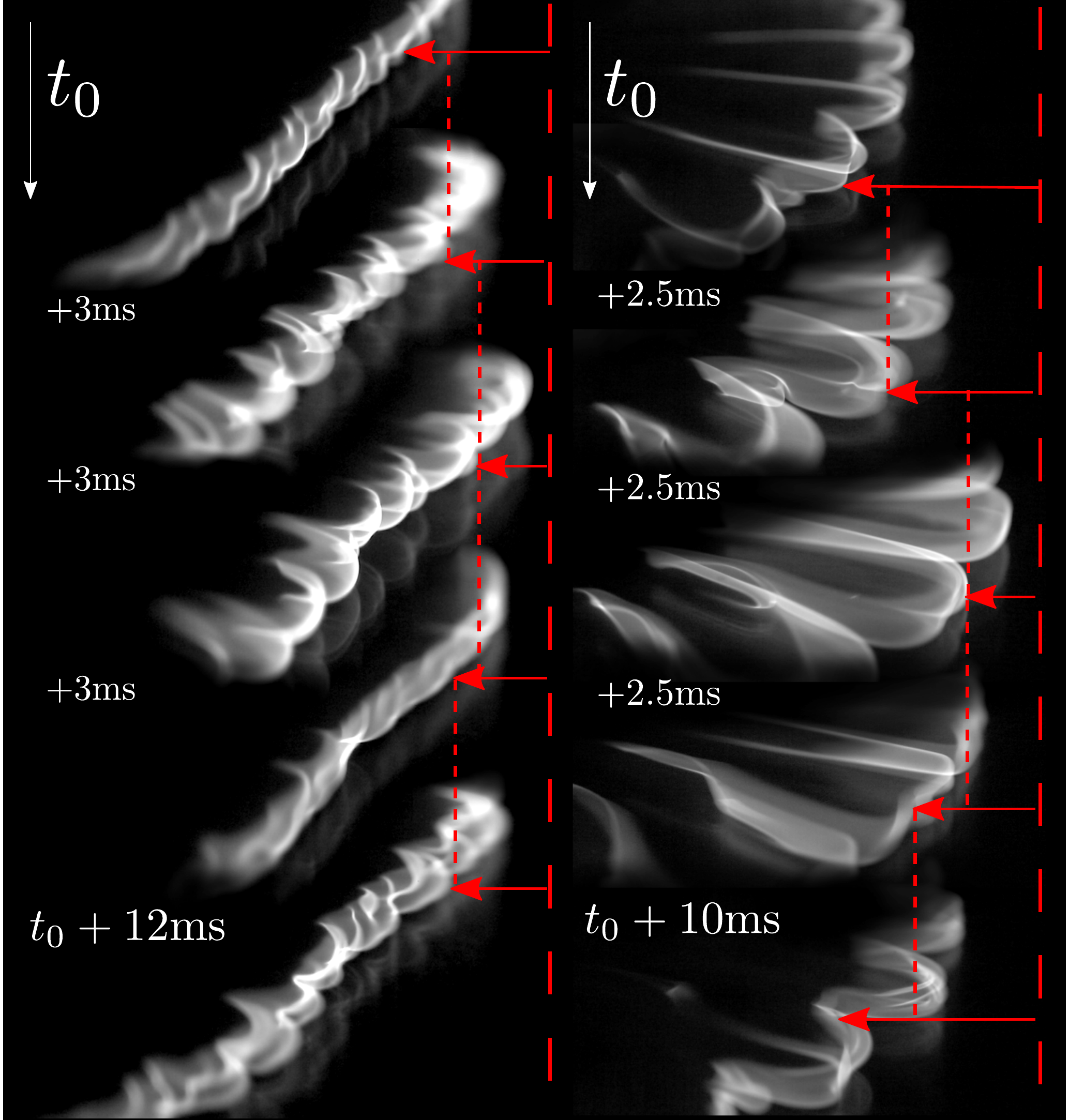}
	\caption{Time sequence illustrating one oscillation cycle of a rich propane flame in the transition stage from the primary to the secondary instability (left images) and once the high-amplitude oscillations are fully developed (right images). Vertical dashed lines establish a common reference for the displacement along each cycle.}
	\label{fig:latcycl}
\end{figure}

To illustrate the change in the flame structure, we show in Figs.~\ref{fig:top}, \ref{fig:lateral} and \ref{fig:latcycl} the top and lateral views of the flame luminous emissions recorded by the camera for lean ($\phi=0.8<\phi_c$) and rich ($\phi=1.1>\phi_c$) propane-air flames propagating, from left to right, towards the closed end of the chamber. The superposed images correspond to the times marked with the respective numbers in Fig.~\ref{fig:composed}a. \\

As it can be seen in Fig.~\ref{fig:top} (top), 70 ms after ignition (stage 1) the lean flame ($\phi=0.8$) wrinkles to form medium-size cells as a consequence of the Darrieus-Landau instability. Soon after, at approximately 300 mm from the ignition source, the flame starts a small-amplitude oscillation that lasts until it reaches the end of the chamber. The flame is flattened in the $x-y$ plane by the acoustic waves, undergoing a smooth oscillatory motion thus slowing its propagation rate down from $U_L=1$ m/s to $U_L=0.44$ m/s, still faster than the velocity of a laminar planar flame ($S_L=0.28$ m/s). When oscillating, neither the form of the flame nor the size of the cells change substantially, as inferred also from a side view of the flame displayed in Fig.~\ref{fig:lateral}. The flame-front cusps, as seen from above, form and merge as observed by Almarcha et al. \cite{almarcha2015experimental} in a downward-propagating propane flame with $\phi=0.6$, despite the fact that no oscillations of the front were reported there.\\

The flame-propagation dynamics change drastically in richer mixtures, as can be seen in the right panels of Fig.~\ref{fig:composed} for a rich propane flame $\phi=1.1$. During the first instants of the process, 40 ms after ignition (stage 1), smaller cells than for the lean flames are formed on the reaction surface and it undergoes an oscillation of small amplitude, characteristic of the first thermo-acoustic instability, that flattens the flame before reaching the first quarter of the chamber. Right after, when the reactive front progresses towards the half of the chamber, the oscillations grow rapidly in amplitude and frequency, accelerating the flame which adopts a marked cellular finger-like shape as observed at stage 4 in Figs.~\ref{fig:top}, \ref{fig:lateral} and \ref{fig:latcycl}.\\

Deeper understanding is achieved by observing the shape of the flame in the transverse direction to the flame propagation (coordinate $z$). To do so, we placed the camera laterally under a certain angle, as described in Fig.~\ref{fig:sketch}. The side views of both lean and rich flames are included in Figs.~\ref{fig:lateral} and \ref{fig:latcycl}. In these flames, the Darrieus-Landau instability induces the formation of cells, smaller as the mixture gets richer, that wrinkle the flame in the $x$-$y$ plane. A smooth parabolic shape in the transversal z-direction, convex towards the fresh mixture, is kept in the early stages of the propagation for both lean and rich mixtures, as can be appreciated in Fig.~\ref{fig:lateral}. Afterwards, for the lean flame, the same parabolic shape is conserved all along its way. On the other hand, the rich flame flattens in both transverse and span-wise directions as the flame reaches 350 mm of the chamber length (stage $3$), consequence of the interaction of the flame with the acoustic waves \cite{aldredge2004experimental,searby1992acoustic}. More detailed photographs of the flame shape during an oscillation cycle performed at this stage are displayed in Fig.~\ref{fig:latcycl} (left). In this figure, we see that the flame front becomes a thin reaction region, nearly confined in the $y$-$z$ plane, and shows small bulges on its surface as a consequence of a wrinkling instability that seems to enhance the oscillatory motion. In the next frames of this picture, the size of the corrugations increases and the flame accelerates building up the pressure in the confined fresh gases before retreating to a new cycle start in the form of an almost-planar wrinkled surface.\\

Returning to Fig.~\ref{fig:lateral}, we can track more clearly how the small-sized bulges, formed on stage $3$, grow across the flame surface, deforming it until it adopts a finger-shaped structure (stage $4$). Later on, the flame initiates the high-amplitude oscillations that extend the reaction front further towards the high-temperature gas region (stages $4$ and $6$). Again, we show in Fig.~\ref{fig:latcycl} (right) a sequence of detailed photographs of the flame during an oscillation cycle once the finger-shaped structure has been developed. During the oscillation, the portion of the flame located at the foremost position (indicated by arrows in the figure) gets delayed forming a cusp pointing towards the burned gas during the drawing back of the flame, behavior that was only reported numerically by Gonzalez \cite{gonzalez1996acoustic}. The violent oscillations continue until the flame reaches the end of the chamber, where small-amplitude vibrations are recovered.\\

\subsubsection{Oscillation frequency analysis} \label{sec:freq}
\begin{figure}[!ht]
	\centering
	\includegraphics[width=0.495\textwidth]{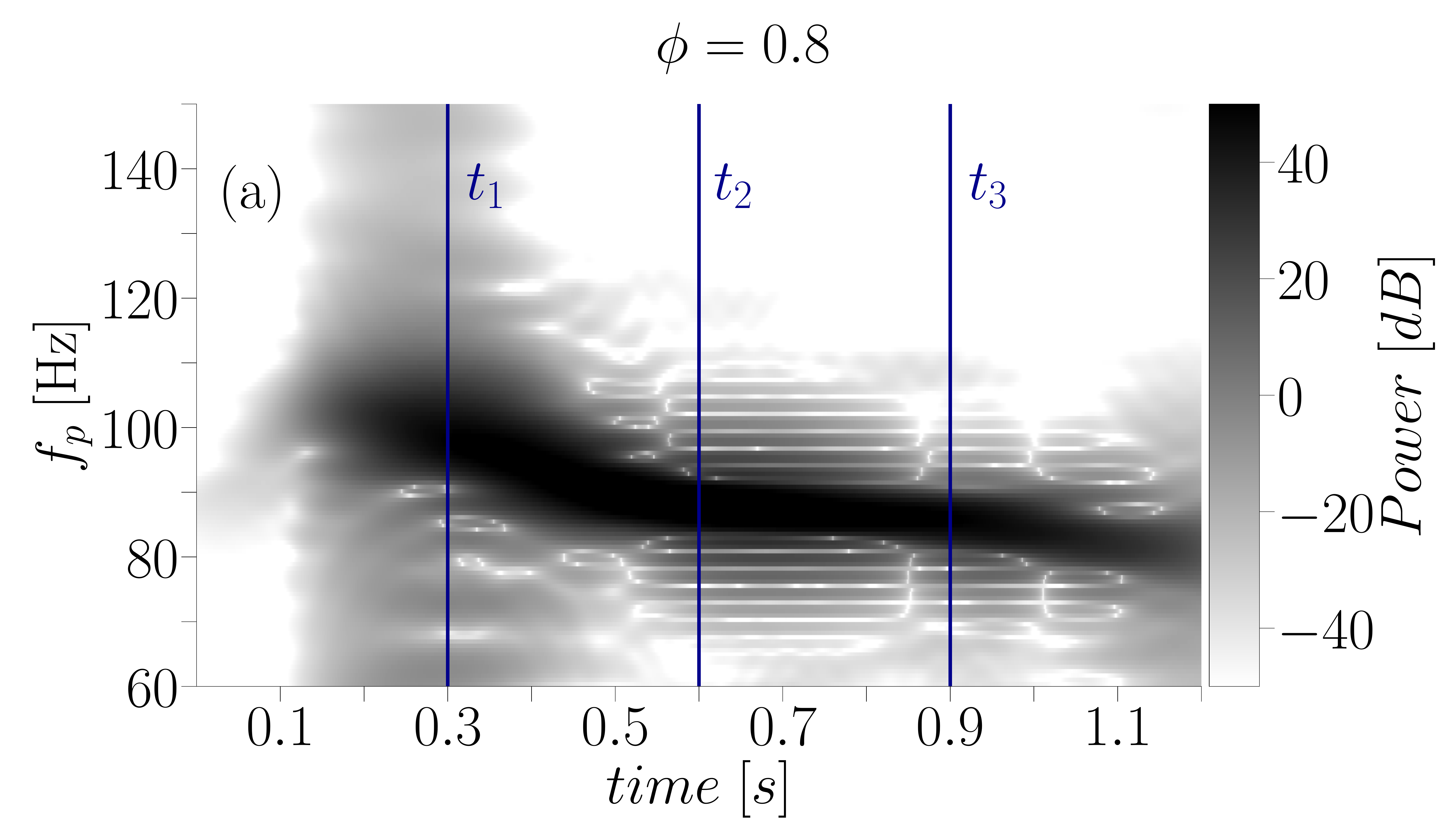}	
	\includegraphics[width=0.495\textwidth]{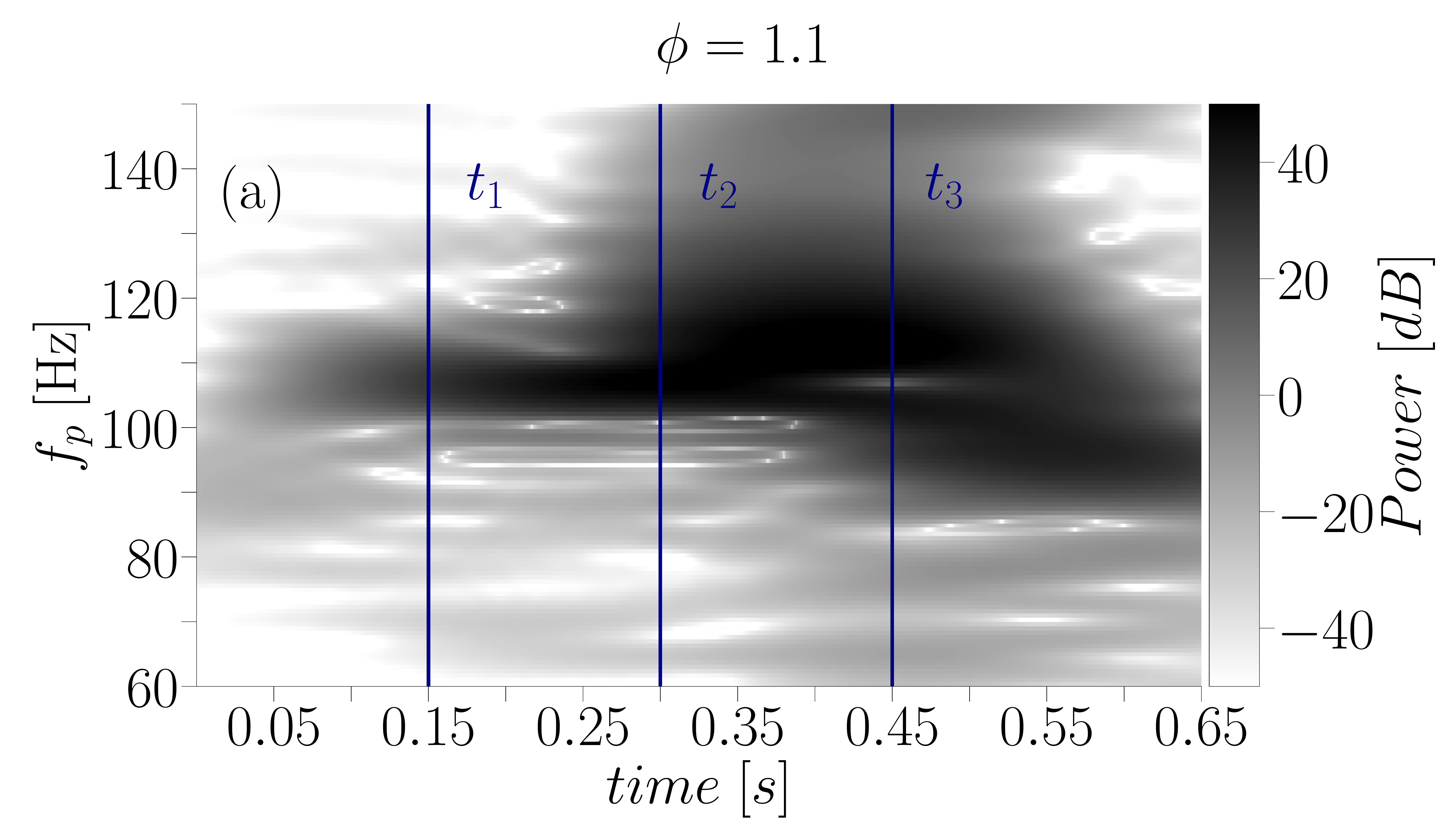}\\	
	\includegraphics[width=0.495\textwidth]{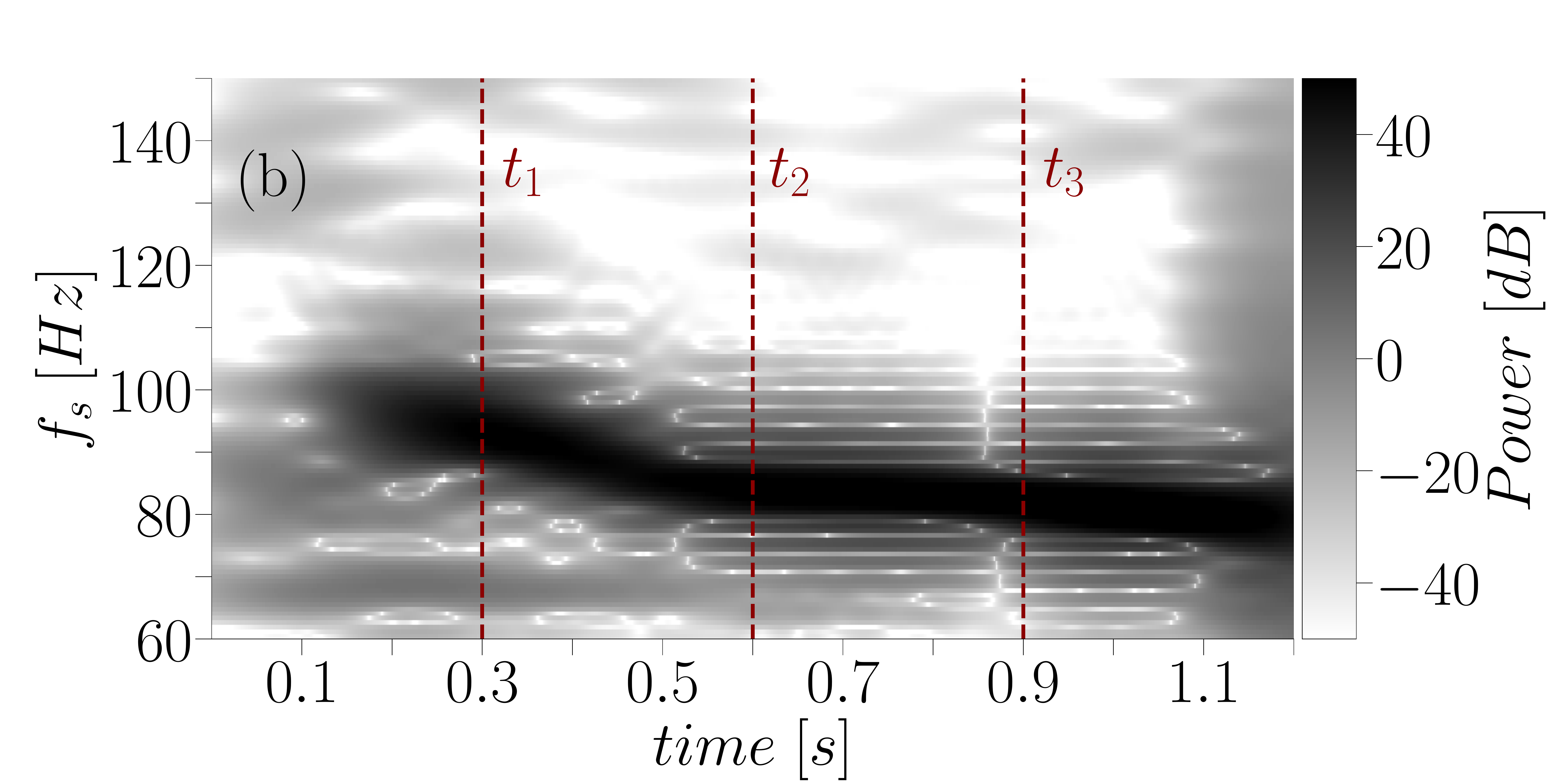}	
	\includegraphics[width=0.495\textwidth]{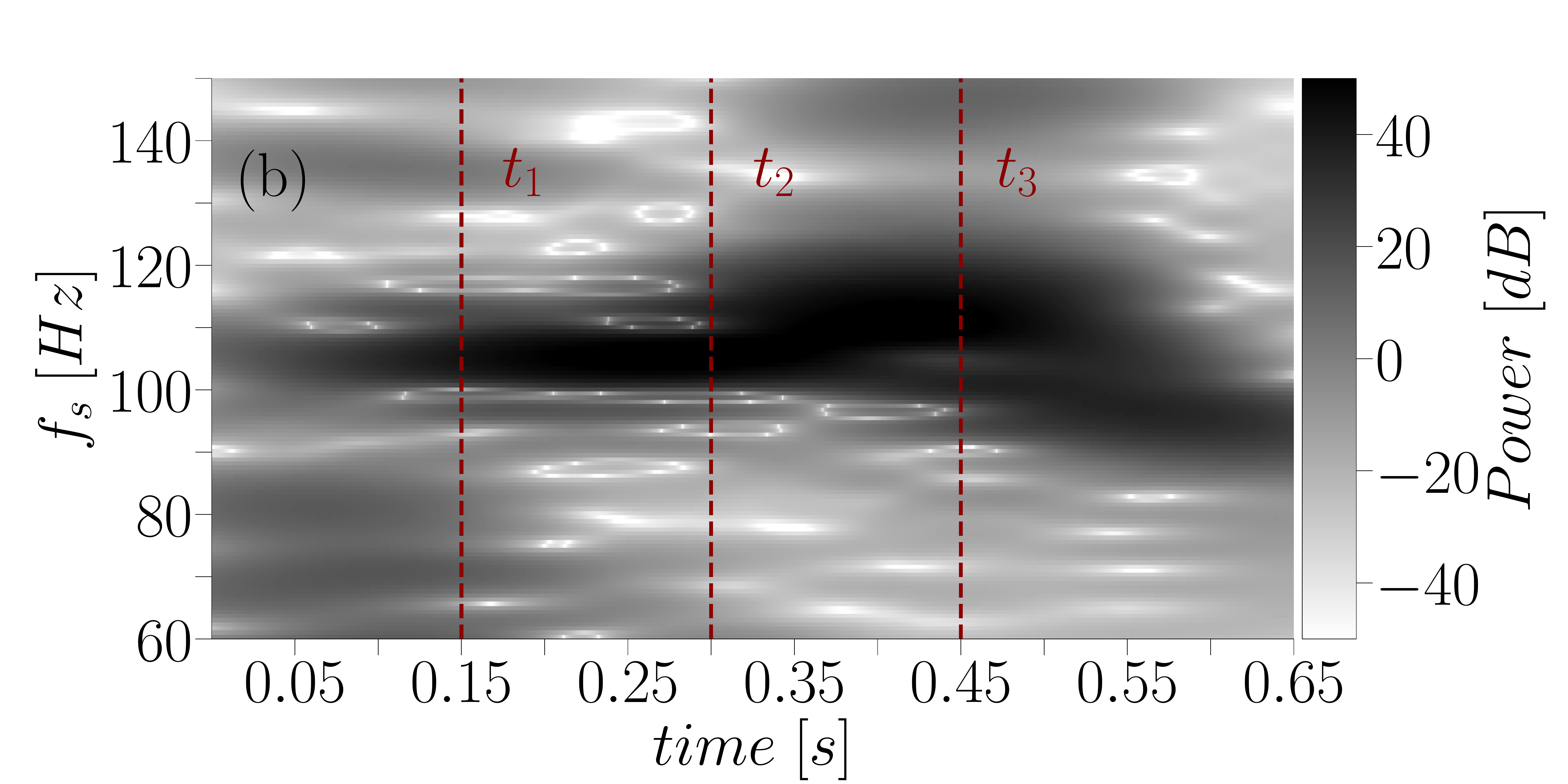}
	\caption{Fourier spectrograms of the  flame position $f_p$ \textbf{(a)} and of the sound level signal $f_s$ \textbf{(b)} for a lean $\phi=0.8$ (left) and rich $\phi=1.1$ (right) propane-air flame.}
	\label{fig:Fourier_spectrograms}
\end{figure}

The comparison between the burning velocity and the acoustic wave shown in Fig.~\ref{fig:composed} suggests a coupling between the two signals. To further investigate this aspect, we represented the Fourier spectrograms of a lean $\phi=0.8$ (left panels) and a rich $\phi=1.1$ (right panels) propane flame in Fig.~\ref{fig:Fourier_spectrograms}. This figure displays a contour map of the evolution with time of the power level $P=20 \log_{10} \left( {\cal{A}}/\bar{{\cal{A}}} \right) $ [dB] stored in every frequency for an oscillatory signal with an instantaneous amplitude ${\cal{A}}$ and average amplitude $\bar{{\cal{A}}}$. According to the color map chosen, the darker the color the more energy is stored in the corresponding frequency. Figures~\ref{fig:Fourier_spectrograms}a and \ref{fig:Fourier_spectrograms}b show the Fourier spectrograms of the burned-area oscillations of the flame images $f_p$ and of the sound signal $f_s$ respectively.\\

\begin{figure}[!ht]
	\centering
	\includegraphics[width=0.495\textwidth]{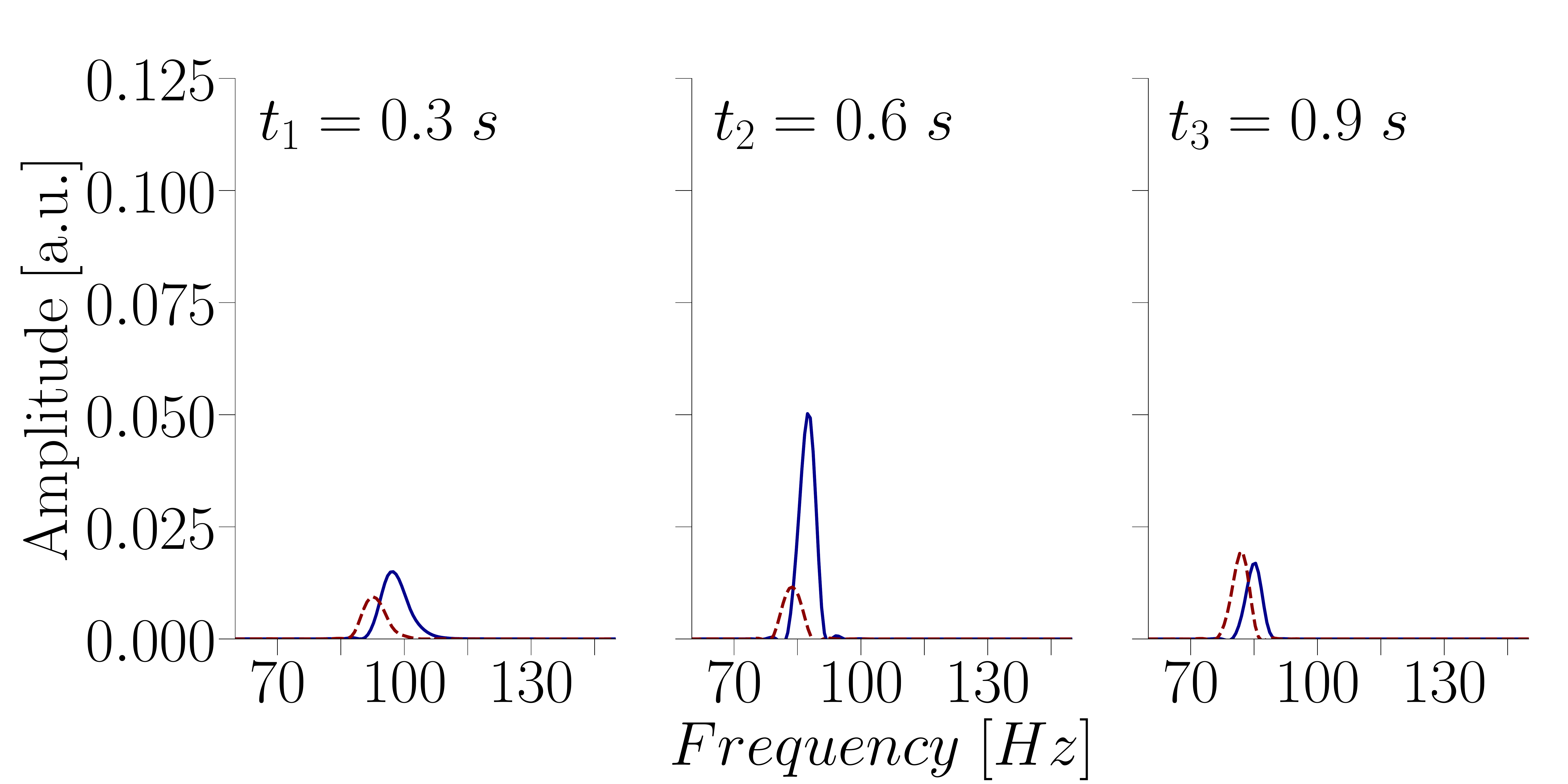}	
	\includegraphics[width=0.495\textwidth]{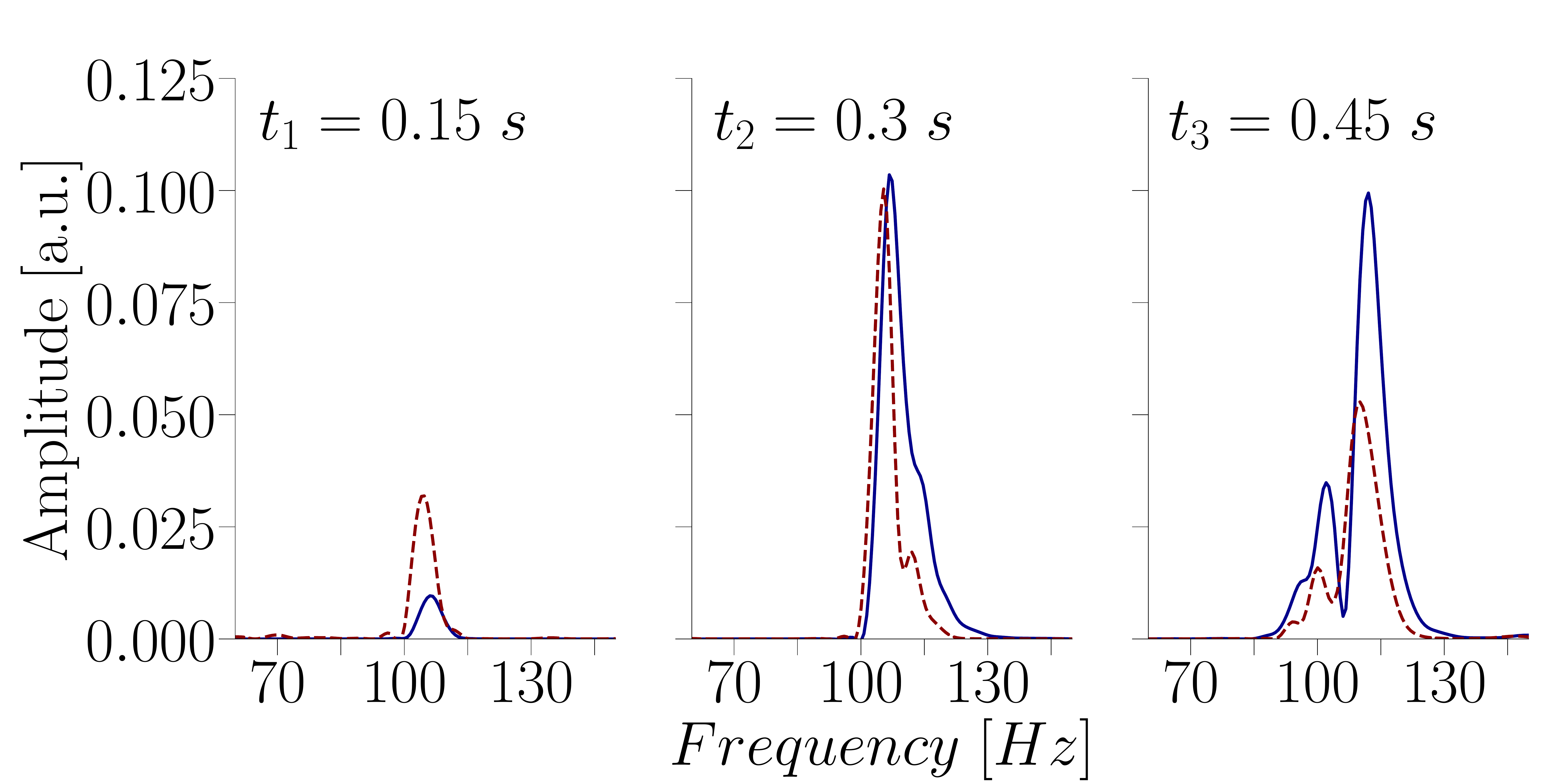}\\	
	\caption{Fourier spectra of the flame position (solid lines) and of the sound level (dashed lines) signals at times $t_1=0.3$ s, $t_2=0.6$ s and $t_3=0.9$ s for the lean flame and $t_1=0.15$ s, $t_2=0.3$ s and $t_3=0.45$ for the rich flame indicated in Fig. \ref{fig:Fourier_spectrograms}. The error of the frequency measurements is $\pm 3$ Hz.}
	\label{fig:Fourier_spectrum}
\end{figure}

For lean propane flames ($\phi=0.8$), small-amplitude oscillations appear at $t_1 \simeq 0.2$~s when the reaction front reaches 200 mm of the total chamber length, at a frequency around 100 Hz that coincides with the frequency of the recorded sound. As the propagation continues along the chamber, the frequency of both the flame oscillations and the pressure wave signal reduces continuously to reach a minimum of 80 Hz at $x=L$. In Fig.~\ref{fig:Fourier_spectrum}, we plot the Fourier spectra at $t_1=0.3$~s, $t_2=0.6$~s and $t_3=0.9$~s comparing the flame-position oscillation and the sound level, where the peak amplitudes match the same frequencies.\\

As expected, rich propane flames ($\phi=1.1$) oscillate with small amplitudes at a frequency around 100 Hz, until the flame-front arrives at the half of the chamber. At this time, the secondary instability emerges and the flame responds undergoing large-amplitude oscillations at frequencies ranging from 100 to 115 Hz. Towards the end of the chamber, the motion is smoothed and the frequency reduces to near 90 Hz. As for lean flames, the Fourier spectra plotted at $t_1=0.15$ s, $t_2=0.3$ s and $t_3=0.45$ s in Fig.~\ref{fig:Fourier_spectrum} show the peak amplitudes of the flame oscillation and of the sound level at the same frequency.

\subsection{Methane (CH$_4$) flames}
\begin{figure}[!ht]
	\centering
	\includegraphics[width=0.495\textwidth]{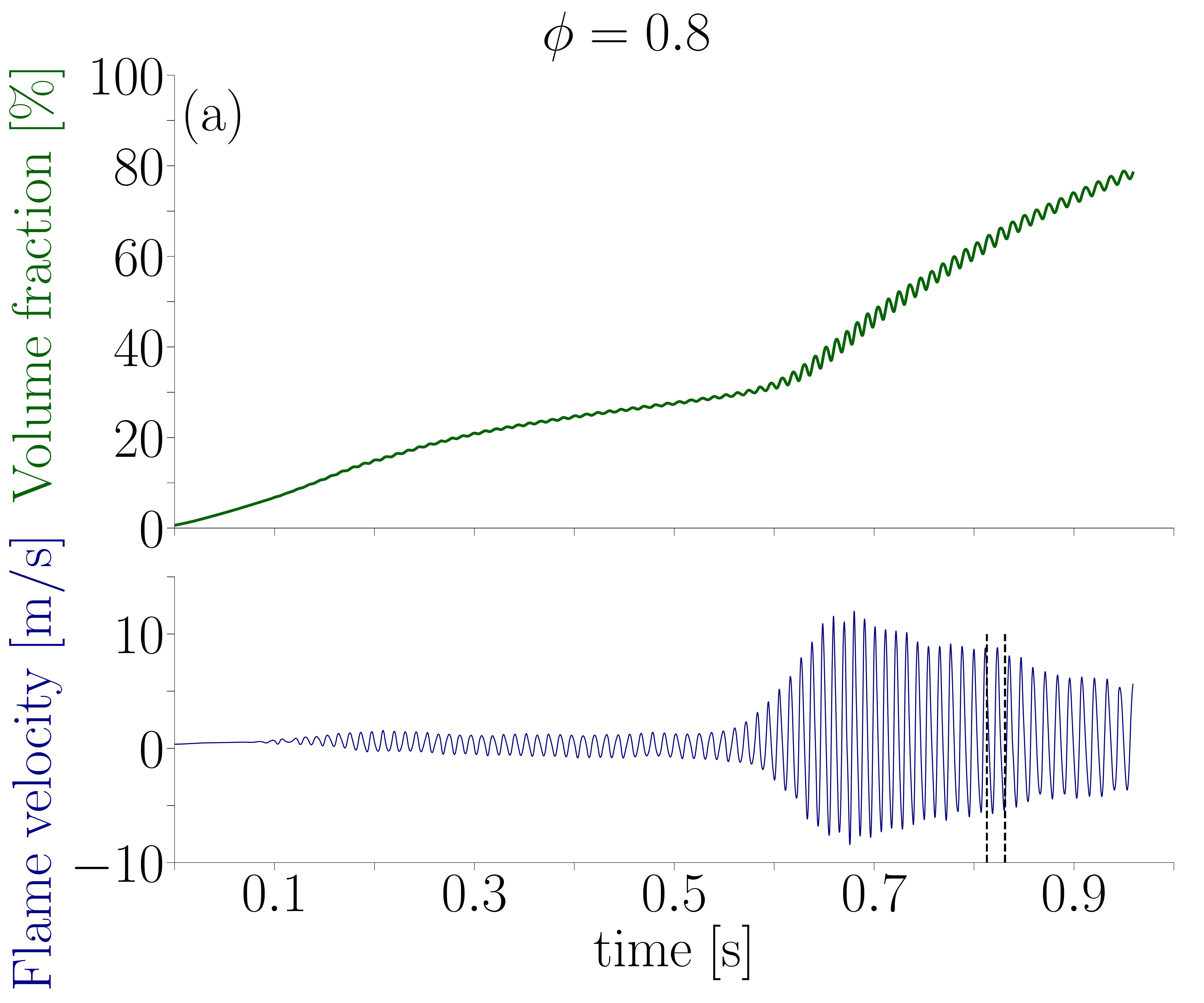} 
	\includegraphics[width=0.495\textwidth]{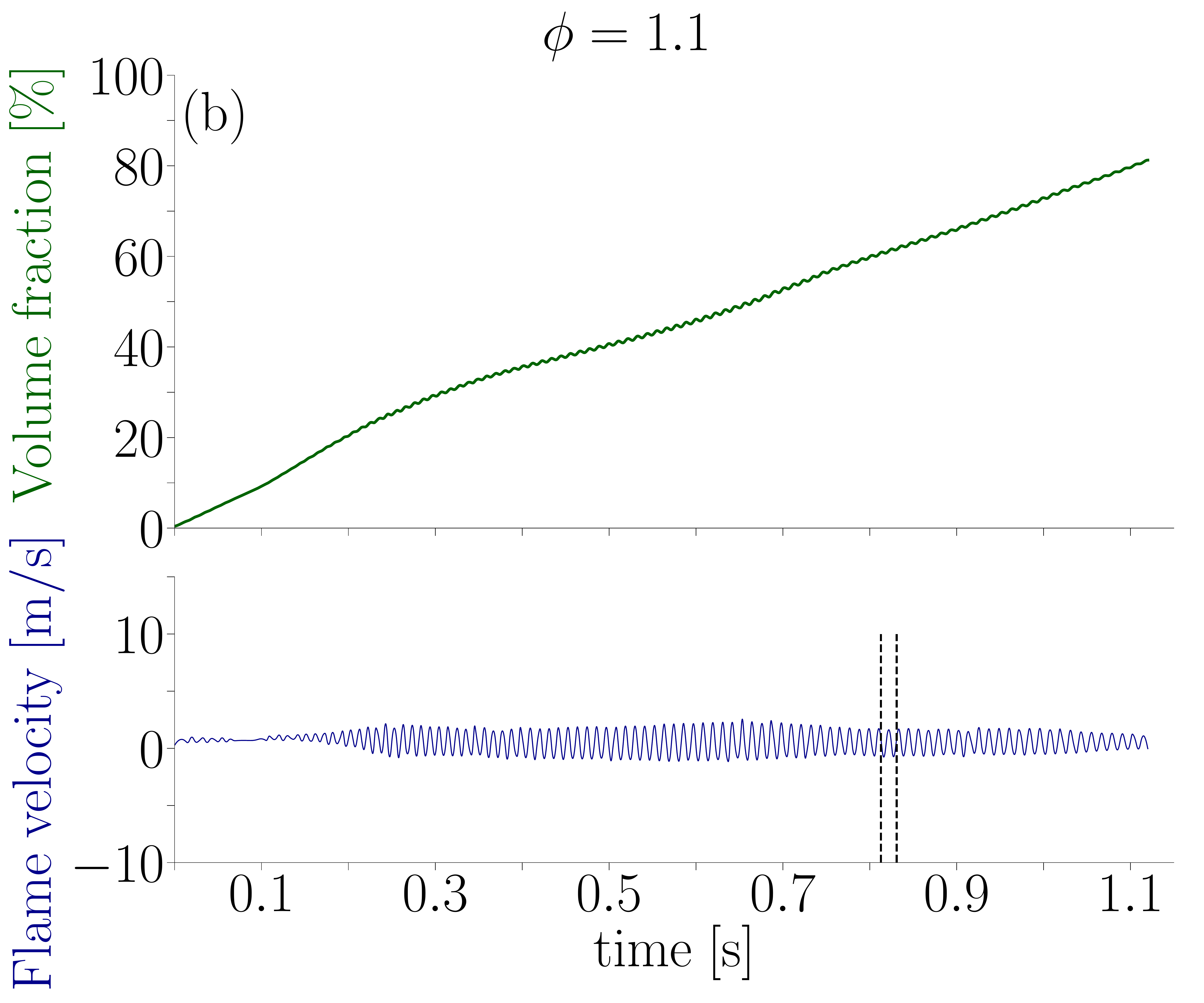}	
	\caption{Time evolution of the burned volume fraction $V_b/V$ (upper thick-solid line) and flame velocity $U_L=(Hh)^{-1} dV_b/dt$ (lower thin-solid lines) for a lean ($\phi = 0.8$) \textbf{(a)} and a rich ($\phi = 1.1$) \textbf{(b)} methane-air flame. The maximum uncertainty of the burned volume measurements is $\pm 2.75\%$. The dotted vertical lines represent the period of oscillation covered by the sequence of images displayed in Fig. \ref{fig:top_methane}}.
	\label{fig:methane_phi08}
\end{figure}

Contrary to propane flames, the secondary acoustic instability is observed in lean methane flames ($\phi<\phi_c \simeq 0.95 \pm 0.05$), as can be checked in Fig.~\ref{fig:methane_phi08}. Lean (rich) methane flames exhibited flame oscillations of similar characteristics to those described above for rich (lean) propane flames.\\

The evolution of both the burned volume fraction and the flame velocity with time is shown in Fig.~\ref{fig:methane_phi08}a and~\ref{fig:methane_phi08}b for lean and rich flames respectively. The oscillation frequency varies with the equivalence ratio and also changes during the propagation of the flame for all the studied cases, although it always remains around 100 Hz. Also, the frequency analysis of the sound generated during the propagation shows a matching with the flame-position oscillation similar to that of propane. To avoid repetition, we do not include in this work the methane equivalent of Figs.~\ref{fig:Fourier_spectrograms} and \ref{fig:Fourier_spectrum}.\\

\begin{figure}[!ht]
	\centering
	\includegraphics[width=0.95\textwidth]{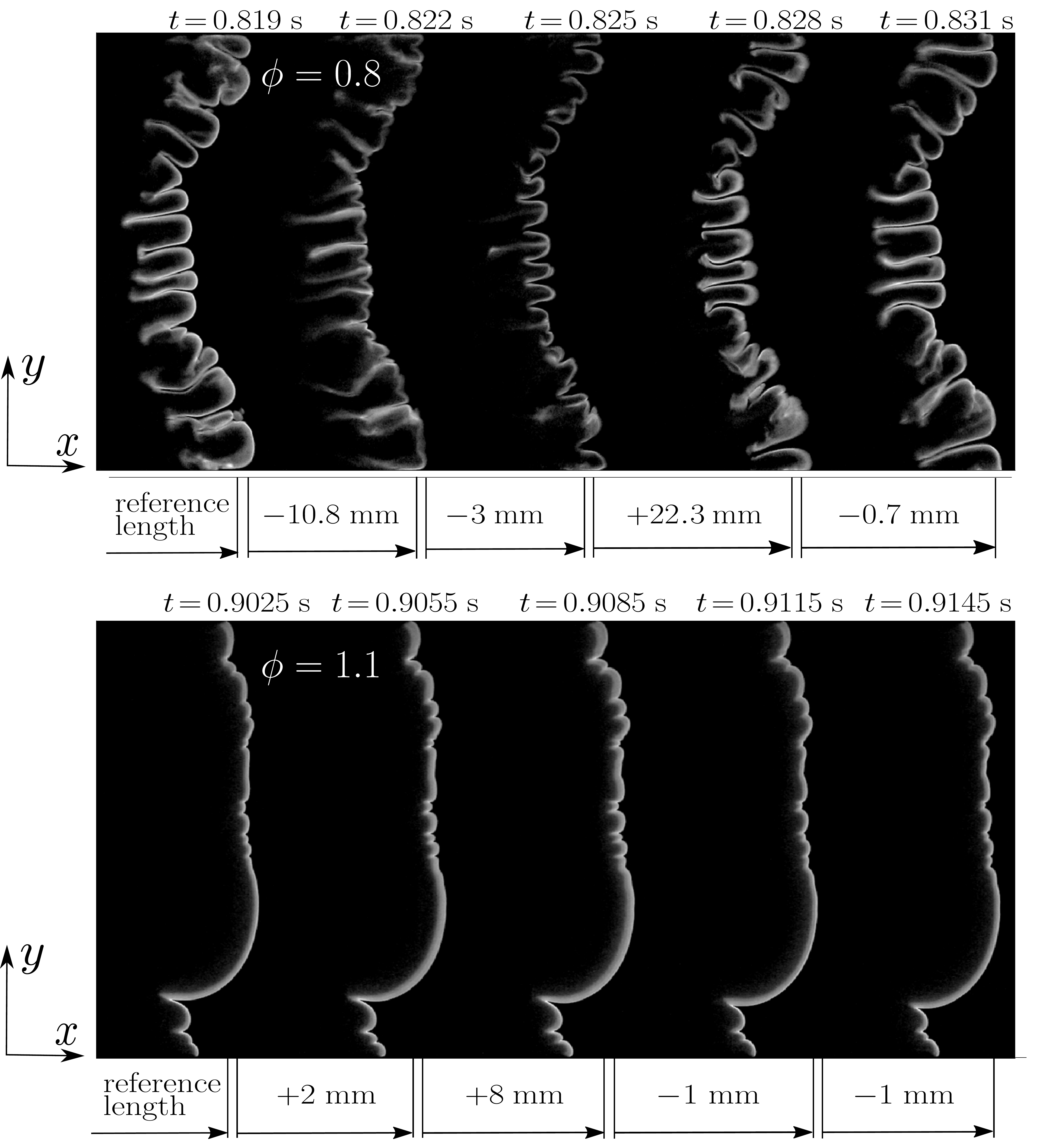}		
	\caption{Top view of methane flames during the oscillation cycle indicated by the vertical dashed lines included in Fig. \ref{fig:methane_phi08}. The leftmost image of the sequence was taken at $t=0.819$ for $\phi=0.8$ and at $t=0.9025$ s for $\phi=1.1$ with the frames separated $\Delta t=0.003$ s. At the bottom of the figure we include the relative distance traveled by the flame between two consecutive frames. The error in the determination of the equivalence ratio is $\pm 0.05$ and the maximum uncertainty of the burned volume measurements is $\pm 2.75\%$}
	\label{fig:top_methane}
\end{figure}

In Fig.~\ref{fig:top_methane} we composed a sequence of images that tracks a lean methane flame ($\phi=0.8$) during an oscillation cycle in which the flame travels from $x\simeq 0.47$~m to $x\simeq 0.48$~m, with the leftmost picture taken 0.819~s after ignition and with the photographs shown every 0.003~s. 
To give an idea of the distance covered by the flame during one oscillation cycle, we included the relative distance traveled by the flame between two consecutive frames measured at one half of the chamber height $y=H/2$. At the beginning of the oscillation cycle, the flame moves backwards a total distance of $-13.8$~mm at $2.3$~m/s. At half of the cycle, the flame swaps the propagation direction to travel forward 22.3~mm with a peak velocity of 7.4~m/s. During the last fourth of the cycle, the flame slowly recedes $-0.7$~mm at 0.23~m/s. The average displacement velocity during this cycle is 0.65 m/s, larger than the flame velocity of a planar flame $S_L=0.24$~m/s.\\

When working with methane, the transition from the primary to the secondary instability takes place at an equivalence ratio close to unity. As shown in  Fig.~\ref{fig:top_methane}  for a rich methane-air flame ($\phi = 1.1$), only small-amplitude oscillations, representative of the primary instability regime, are detected. The net displacement along the selected cycle is $8$~mm, slightly shorter than the distance traveled by the lean flame ($\phi=0.8$) under the secondary oscillatory regime, and the average velocity at this stage of propagation is 0.67 m/s. Regarding the shape of the flame-front, only minor changes can be appreciated when experiencing these small-amplitude oscillations, as shown in Fig.~\ref{fig:top_methane}. \\

\subsection{DME (CH$_3$OCH$_3$) flames}
\begin{figure}[!ht]
	\centering
	\includegraphics[width=0.495\textwidth]{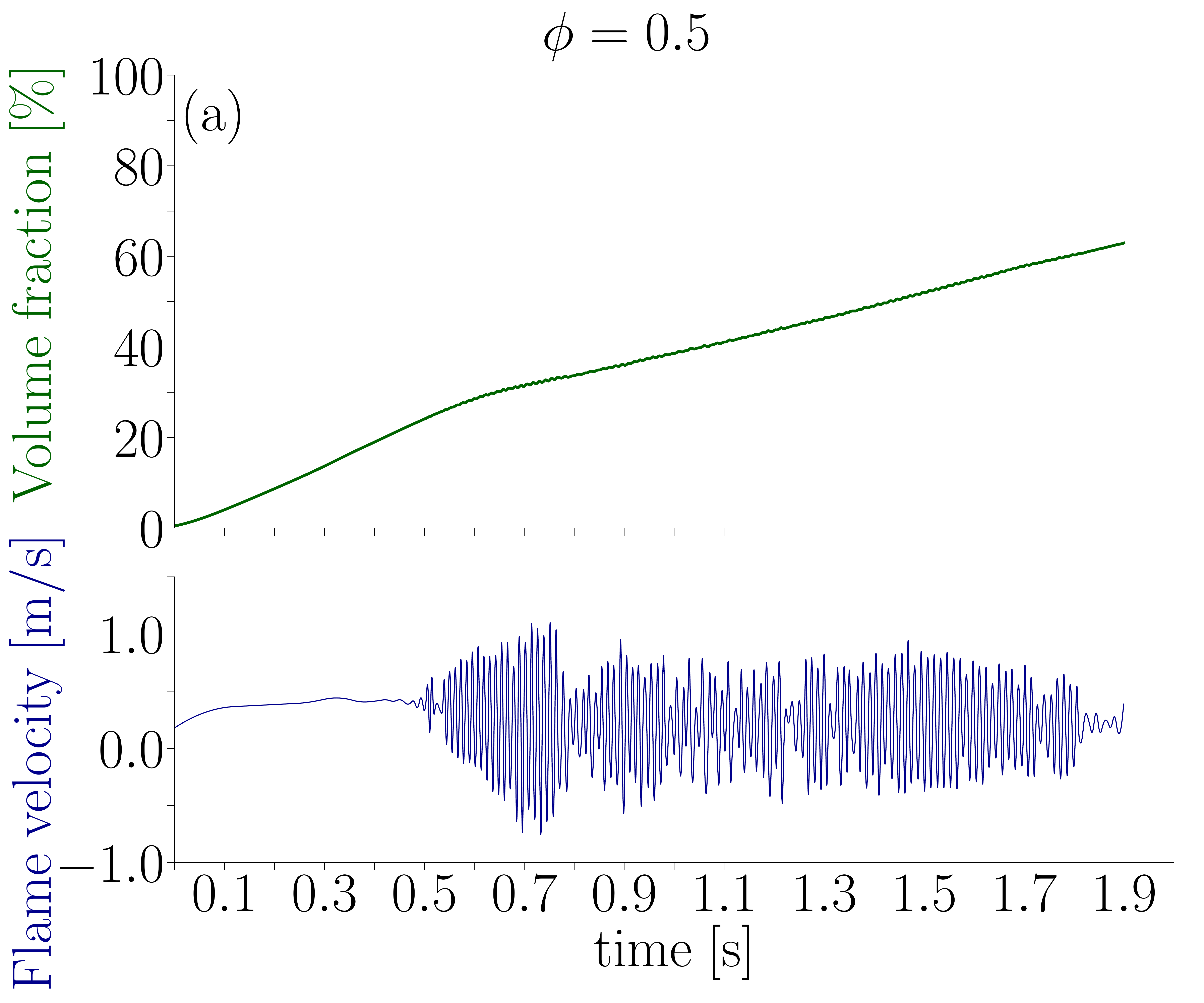}
	\includegraphics[width=0.495\textwidth]{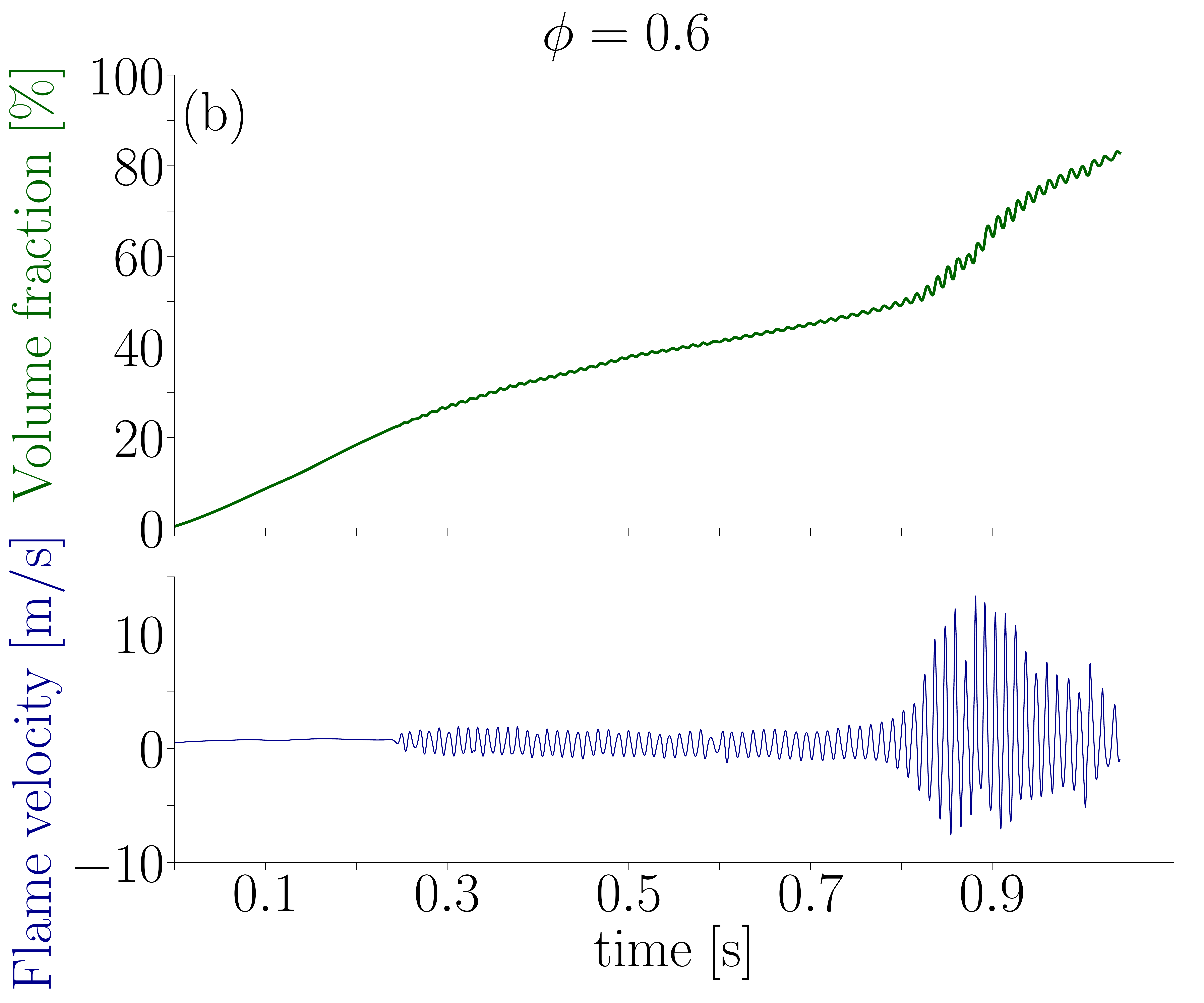}
	\caption{Time evolution of the burned volume fraction $V_b/V$ (upper thick-solid line) and the flame velocity $U_L=(Hh)^{-1} dV_b/dt$ (lower thin-solid lines) for an equivalence ratio $\phi =0.5 < \phi_c$ \textbf{(a)} and $\phi = 0.6>\phi_c$ \textbf{(b)} of a DME-air flame. The maximum uncertainty of the burned volume fraction measurements is $\pm 2.75\%$. The range of the vertical axes is different for the flame velocity plots.}
	\label{fig:DME_pos_vel}
\end{figure}

As for propane, the secondary instability has been observed  only for equivalence ratios above a critical value that turns out to be  approximately $\phi_c \simeq 0.55 \pm 0.05$. As illustrated in Fig. \ref{fig:DME_pos_vel}, the evolution of the burned volume fraction and of the flame velocity with time for $\phi=0.5<\phi_c$ and $\phi=0.6>\phi_c$ shows the previously-observed characteristics of propane and methane flames oscillating in the primary and secondary acoustic instability regimes, respectively. Note that the vertical scales of the flame velocity were modified for clarity. It is only in the case $\phi=0.6>\phi_c$ when the finger-shaped, large-amplitude flame oscillations are observed, with maximum and minimum oscillation velocities near 13 and -7 m/s respectively.\\

To achieve the combustion of flames as lean as $\phi=0.5$ we did not preheat the gas nor the cell walls, and the experimental procedure was identical to that described above in section \ref{sec:experimental_procedure}. Such lean flames showed considerably longer propagation times, of around 2 seconds, and the post processing of the images was significantly harder due to the low emissivity of this flames, making Fig.~\ref{fig:DME_pos_vel} look noisier than the figures for propane and methane.\\

\section{The effect of the combustion chamber thickness $h$}
\subsection{Heat losses}
In this subsection, we give an order-of-magnitude analysis comparing the characteristic values of the different heat-loss mechanisms that might affect the flame propagation under the prescribed conditions, an important point that has been recurrently disregarded in previous studies.\\

First, the papers by Searby \cite{searby1992acoustic}, Ya\~nez et al. \cite{yanez2015flame}, Almarcha et al. \cite{almarcha2015experimental}, Gross et al. \cite{Gross2014} and Sharif et al. \cite{sharif1999premixed} did not mention the influence of heat losses on their results, while Aldredge and Killingsworth \cite{aldredge2004experimental} simply indicated that their effect was not important. In turn, Yoon et al. \cite{yoon2016onset} only took acoustic losses into account.\\

Considering the worst case scenario, at which the inner faces of the horizontal plates are at room temperature $T_u$, we can estimate the relative importance of the heat losses by comparing the conductive heat losses $q_k \sim H \delta_T k_g (T_b-T_u)/h$ from the flame to the horizontal plates per unit time through an area $H \delta_T$, and the heat released by the flame per unit time $q_f\sim \rho S_L Q Y_u h H$, yielding

\begin{equation}
\Delta=\dfrac{q_k}{q_f} \sim \left(\dfrac{\delta_T}{h}\right)^2, \label{eq:heat_losses}
\end{equation}
where $QY_u=c_p(T_b-T_u)$ is the heat release per unit of fuel mass consumed, $Y_u$ the fuel mass fraction of the unburned gas and $c_p$ and $k_g$ are the air specific heat and the thermal conductivity, respectively. Using the data summarized in Table \ref{table:flame_parameters} to provide the characteristic values for stoichiometric propane-air flames ($\delta_T=0.05$ mm, $h=10$ mm), we obtain typical values of  $\Delta \sim 25 \times 10^{-6} \ll 1$, which confirm the small influence of the heat losses in the widest-channel experiments. As the channel height $h$ was progressively reduced by setting the horizontal plates closer to each other, the effect of the heat losses would become more important, leading to flame extinction for values of $\delta_T/h \sim O(1)$. \\

The heat lost by conduction to the channel's walls, even when small compared with the heat released by the flame, might be conducted longitudinally along the solid wall upstream of the flame and transfered back to the gas, preheating the mixture before the arrival of the reaction front. The time for this to occur in a distance $\delta_T$ upstream of the traveling flame along the solid wall is $t_{kl} \sim \delta_T^2/D_s$, where $D_s$ is the thermal diffusivity of the solid wall. When compared with the residence time of the flame $t_f \sim \delta_T/S_L$, it is possible to neglect the preheating of the gas close to the solid surfaces as long as the ratio $t_{kl}/t_f \sim D_T/D_s \gg 1$. In the case of the glass ($D_s=3.4 \times 10^{-7}$ m$^2$/s) and the vinyl sheet ($D_s=5 \times 10^{-8}$ m$^2$/s) that form both the upper and lower horizontal plates of our experimental setup,  the criterion $t_{kl}/t_f \gg 1$ is satisfied. In experiments that use quartz ($D_s= 1.4 \times 10^{-6} $ m$^2$/s) or metal covers ($D_s \sim 10^{-4} $ m$^2$/s), one should be cautious to properly assess the influence of this effect on their results.\\

\begin{figure}[!ht]
	\centering
	\includegraphics[width=0.495\textwidth]{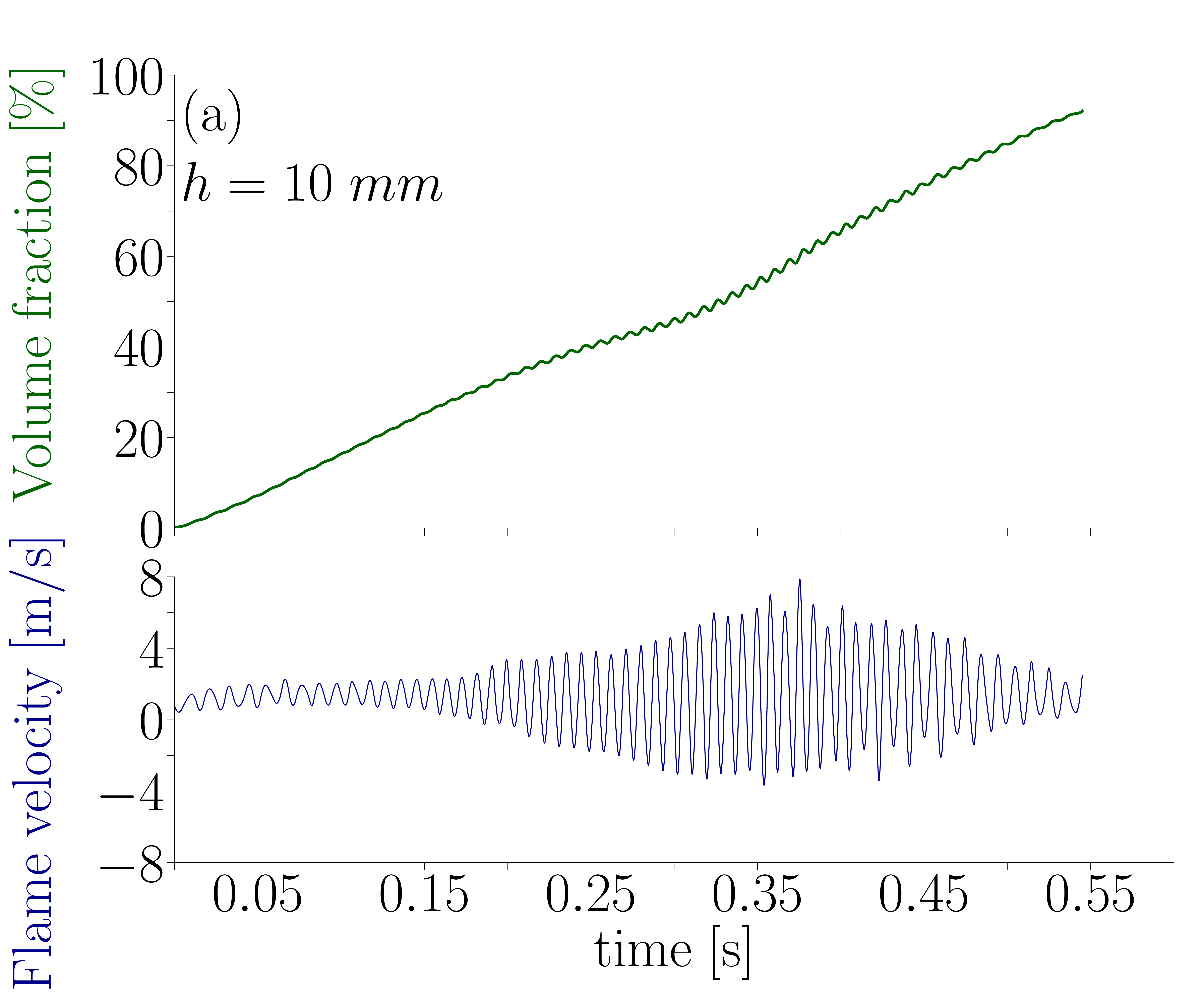}
	\includegraphics[width=0.495\textwidth]{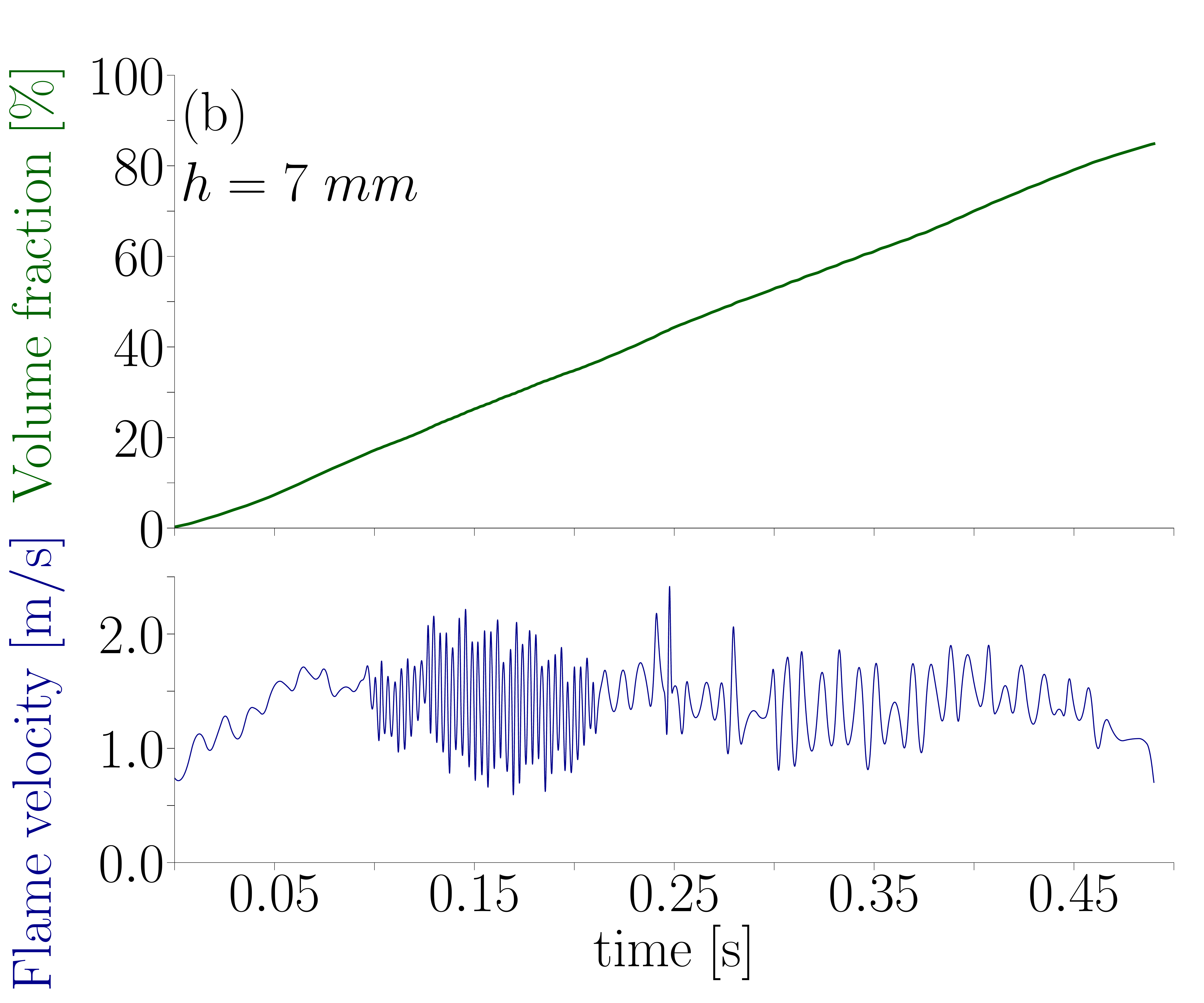} \\
	\includegraphics[width=0.495\textwidth]{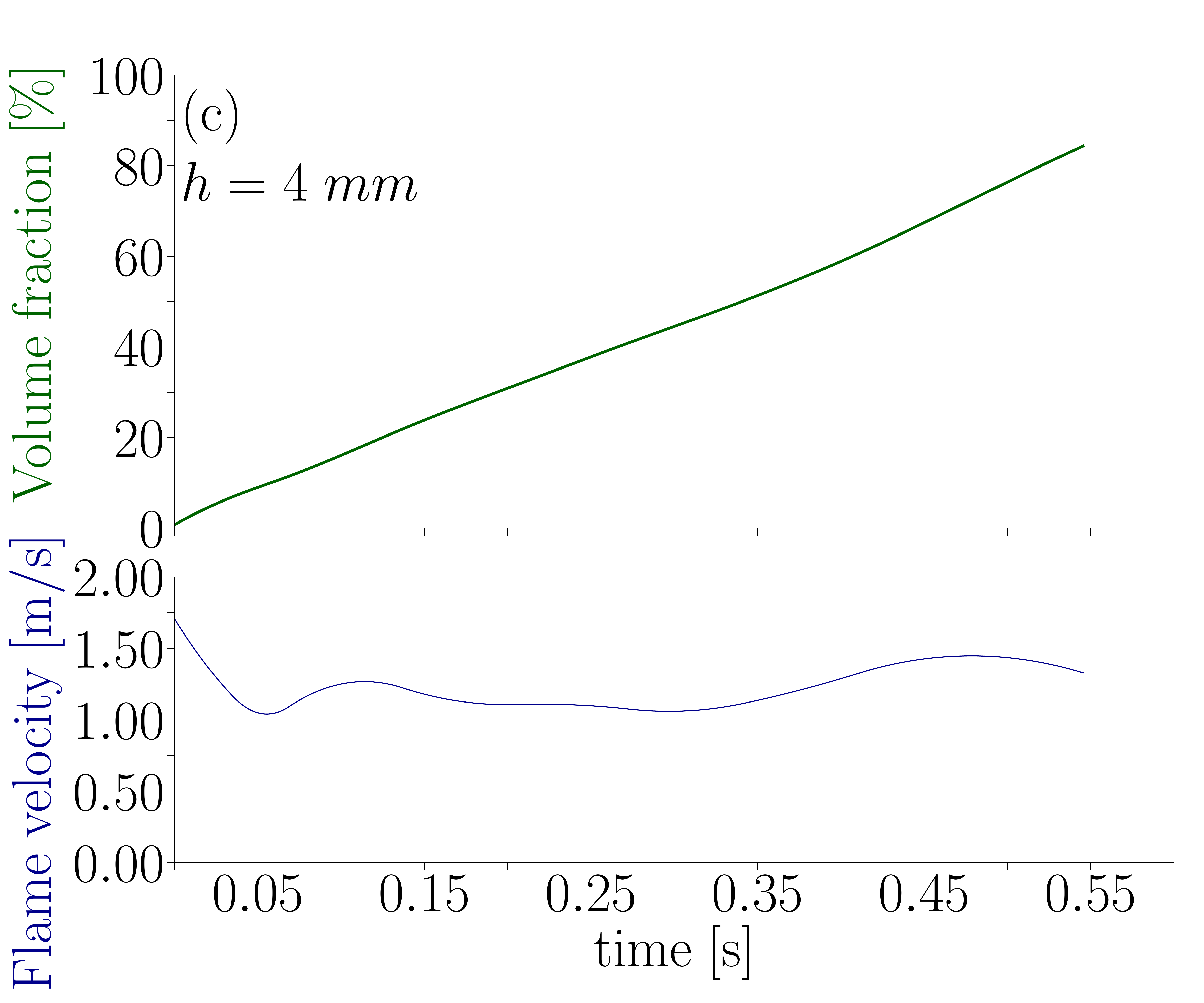}
	\includegraphics[width=0.495\textwidth]{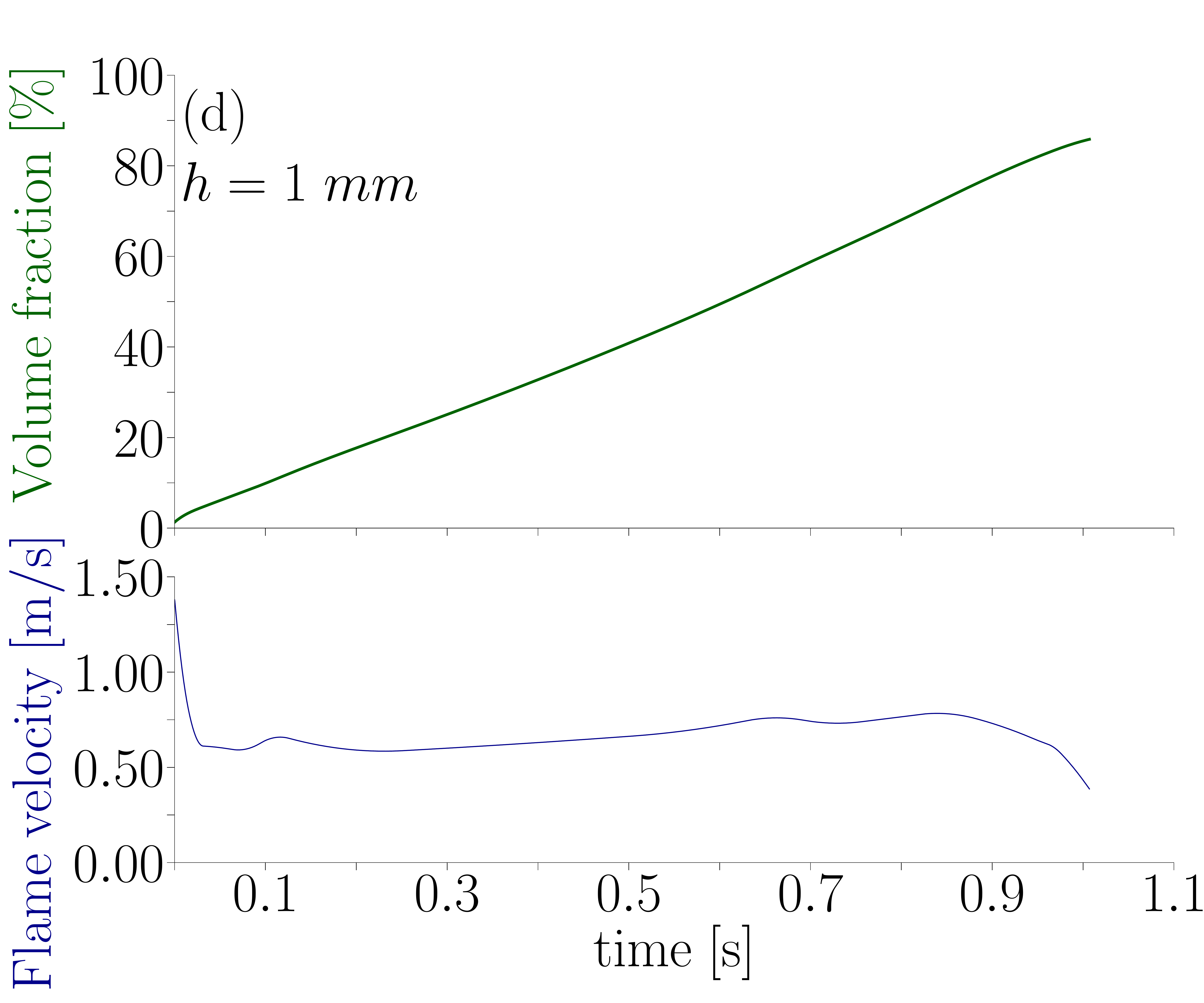} \\
	\caption{Burned volume fraction and flame velocity of a DME flame with $\phi=1$ in a channel of thickness (a) $h=10$ mm (b) $h=7$ mm, (c) $h=4$ mm and (d) $h=1$ mm. The range of the axes is different in each plot. }
	\label{fig:DME_position_h}	
\end{figure}

Most of the heat losses take place in the burned region from the high temperature gas to the walls. Its potential importance on the flame propagation can be estimated by calculating the characteristic temperature gradient downstream of the flame. By considering the heat losses to the walls $k_g (T_b-T_u)/h$ in a reference system attached to the flame, we can estimate the temperature change in a portion of the channel of length $l$ by doing an energy balance in a control volume of height $h$ with the sides against the upper and lower walls to give $(T_b-T_s)/(T_b-T_u)\sim (\delta_T/h)(l/h)$, being $T_s$ the gas temperature at a distance $l$ downstream of the flame and $T_u$ the solid wall temperature. For the flame to be affected by the negative temperature gradient downstream of the flame, the temperature change in a region $l \sim \delta_T$ should be of the order $(T_b-T_s)/(T_b-T_u) \sim (\delta_T/h)^2 \sim \beta^{-1}$ \cite{Zeldovich_1938}, where $\beta$ is the Zeldovich number. Using the data included in table \ref{table:flame_parameters} of the manuscript, we obtain $(\delta_T/h)^2 \ll \beta^{-1}$ in most of the cases tested in our experiments. It is only in very narrow channels $h = 1$ mm when the negative temperature gradient $(\delta_T/h)^2 \sim \beta^{-1}$ seems to affect the propagation of the flame, as shown in Fig. \ref{fig:DME_position_h} for a stoichiometric DME flame. Note that the large values of $\beta$ would make this effect more restrictive than the direct heat losses of the flame to the walls analyzed above to write Eq. \ref{eq:heat_losses}\\

As expected by the estimations given above, the smaller the chamber thickness the longer it takes to the flame to reach the end of the chamber due to the effect of the heat losses as the volume-to-surface ratio is increased. Using Fig.~\ref{fig:DME_position_h} to obtain the average propagation velocity, we can conclude that flames propagating in chambers with $h=10, 7$ and $4$ mm, with approximately the same propagation time, are weakly affected by the heat losses. It is only in the case $h=1$ mm when the deceleration of the flame becomes noticeable, presumably because of the heat losses. Such small effect of $h$ in the flame velocity seems to indicate that heat losses could be considered negligible except in very narrow channels.
These estimations have been examined numerically by Kurdyumov and Fern\'andez-Tarrazo \cite{kurdyumov2002lewis}, Daou and Matalon \cite{daou2002influence} and S\'anchez-Sanz \cite{Sanchez2012} confirming the small influence of heat losses for sufficiently wide channels. \\

\begin{figure}[!ht]
	\centering
	\includegraphics[width=0.95\textwidth]{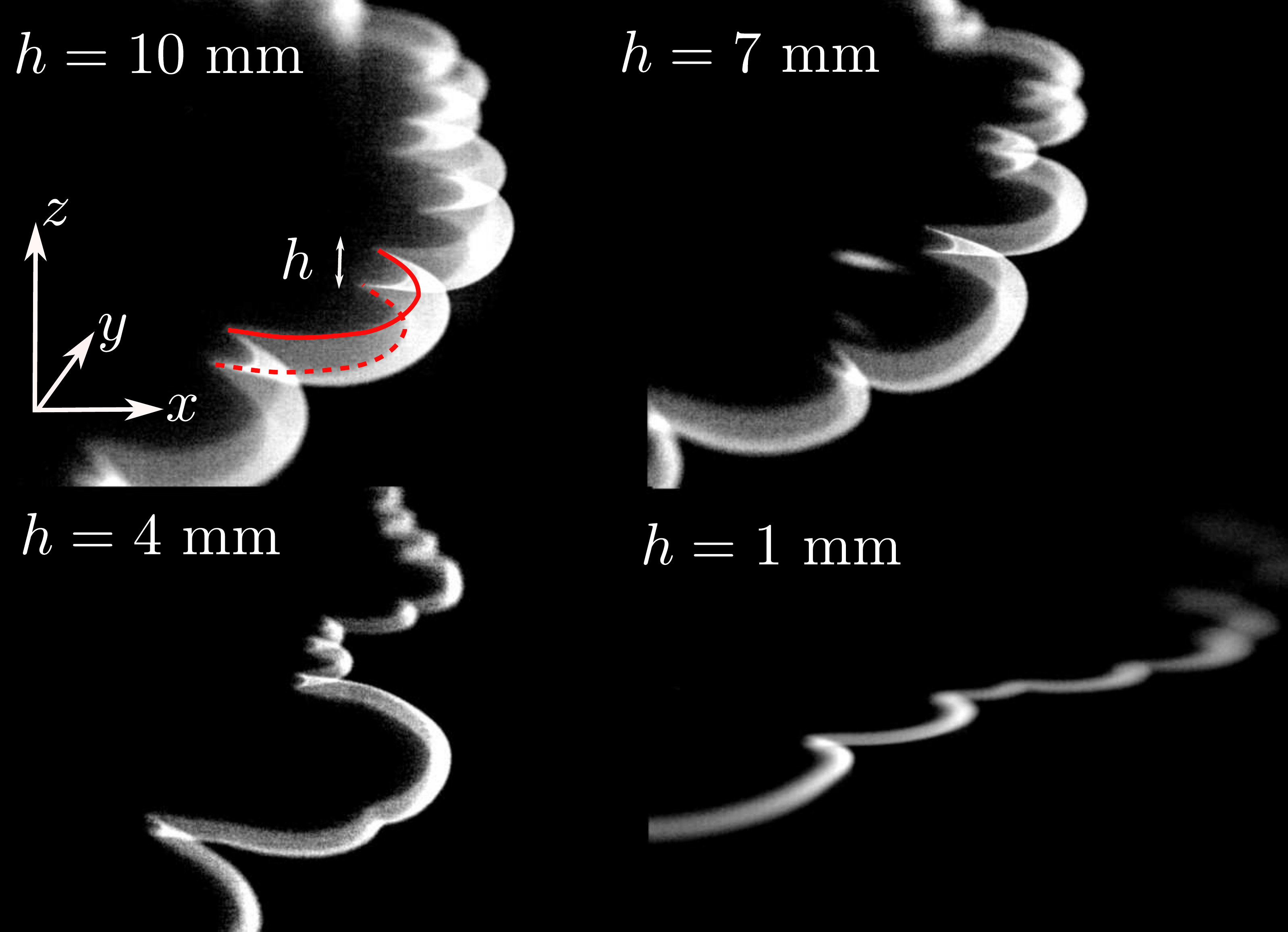}	
	\caption{Lateral view of a stoichiometric DME flame propagating in a  $h=10$ mm, $h=7$ mm, $h=4$ mm and $h=1$ mm cell. The solid and dashed red lines indicate where the flame touches the upper and lower horizontal plates, respectively  }\label{fig:DME_lateral}	
\end{figure}

Previous numerical studies \cite{Galisteo2018,Kurdyumov2016,kang2006effects} reduced the computational cost of their computations by reducing the number of dimensions of the problem by assuming that, in the limit of very narrow channels $h/\delta_T \ll 1$, the transverse profiles ($z$ axis) of temperature and mass fraction are uniform. To check this point, we studied the effect of the channel gap on the flame shape in  Fig. \ref{fig:DME_lateral}. In this figure we included the lateral view taken at $x \sim  150$ mm of a stoichiometric DME flame propagating in a chamber with $h=10, 7$, $4$ and $1$ mm. In this figure we can observed how the reaction region keeps its parabolic shape even in very narrow channels $h=1$ mm in which the parameter $h/\delta_T \sim 20$. Much narrower channels seem to be necessary to reach uniform transverse profiles of temperature and mass fraction.

\subsection{Acoustic losses}
Petchencko et al. \cite{petchenko2006violent,petchenko2007flame} performed a numerical study of a flame propagating towards the close end of a two-dimensional channel. They reported the violent folding of the flame as consequence of a flame acoustic resonance. According to the authors, as the ratio $a=h/\delta_T$ was reduced, the amplitude of the flame oscillation decreased but never disappeared. To test their results, we modified the chamber thickness by stacking 3 mm-thick PVC laminae to progressively reduce the gap between the horizontal plates from $h=10$ mm to $h=1$ mm. The evolution of the burned volume fraction and the flame velocity with time for $h=10, 7, 4$ and $1$ mm are plotted in Fig.~\ref{fig:DME_position_h} for DME with $\phi=1$.\\

In the widest chamber ($h=10 $ mm) the flame front presents high-amplitude oscillations similar to that shown above in Figs.~\ref{fig:composed}, \ref{fig:methane_phi08} and \ref{fig:DME_pos_vel} for propane, methane and DME respectively. The flame travels along the chamber with a flame speed that oscillates between a maximum of 8 m/s and a minimum of -5 m/s approximately. The high-amplitude oscillations were not observed when the chamber thickness was reduced to $h=7$ mm. Both the burned volume fraction and the flame velocity display small oscillations that are characteristic of the primary acoustic instability regime. Farther reduction of the chamber height ($h=4$ mm and $h=1$ mm) completely eliminates the flame oscillations and the flame propagates with a non-oscillatory velocity along the chamber.\\

To understand the effect of the channel height on the flame oscillations, it is worthwhile to estimate the order of magnitude of the acoustic energy and of the acoustic viscous dissipation. When an harmonic sound wave, with angular frequency $\omega$, propagates along a channel, an acoustic boundary layer of thickness $\delta_v \sim (2 \nu /\omega)^{1/2}$, being $\nu$ the kinematic viscosity, appears near to the channel walls to accommodate the fluid particle oscillation amplitude from its value in the mainstream to zero on the wall's surface. At a sound frequency of around 100 Hz measured in our experiments, the acoustic boundary layer is of order $\delta_v \sim 0.2$ mm in the cold gas  ($T_u=273$ K, $\nu=15.06 \times 10^{-6}$ m$^2$/s) and grows to be as thick as $\delta_v \sim 1.12$ mm in the burned gas ($T_b=2000$ K, $\nu=39.43 \times 10^{-5}$ m$^2$/s). Such a thick acoustic boundary layer brings associated a strong dispersion of acoustic energy that can be estimated easily by comparing the residence time of a sound wave $t_a \sim 2L/c$, with $c\simeq 695$ m/s representing the sound velocity at a characteristic temperature $T \sim (T_b+T_u)/2 \sim 1000$ K, and the acoustic dissipation time $t_d \sim \rho h^2/\mu$, calculated by comparing the energy of the acoustic sound wave per unit volume $\rho v^2$, and the viscous dissipation rate $\mu (v^2/h^2)$, with $v$ being the velocity induced by the acoustic pressure wave. Comparing both characteristic times we obtain
\begin{equation}
\dfrac{t_a}{t_d} \sim \dfrac{2 \nu L}{c h^2} \sim  
\left\{\begin{array}{cc}
0.005 & \textrm{   for   } h=10 \textrm{ mm}. \\ 
0.497 & \textrm{   for   } h=1 \textrm{ mm}. \\ 
\end{array}
\right.
\end{equation}
Therefore, for $h=10$ mm, the acoustic time is much shorter than the acoustic dissipation time, what indicates that the pressure wave has time enough to interact with the flame before its attenuation. Contrarily, in smaller chamber gaps, the dissipation time is of the order of the acoustic time, limiting the effect of the acoustic wave on the flame. The parametric dependence of the ratio $t_a/t_d$ coincides with the dissipation rate due to viscosity $(\nu L /c h^2)^{1/2}$ of a pressure wave traveling along a channel of height $h$ calculated by Blackstock \cite{blackstock2000fundamentals} and by Clanet et al. \cite{clanet1999primary}. Notice that the characteristic time for radiative damping through the open end $t_{rad} \sim c L/(\omega h)^2$ \cite{clanet1999primary} is much longer than $t_d$ and can therefore be neglected.\\

The transition to the secondary instability has also been observed in narrower channels in our experiments when richer DME-air mixtures were used. For example, for $h=7$ mm and $h=4$ mm, high-amplitude oscillations are only observed for $\phi>\phi_c=1.175 \pm 0.05$ and $\phi>\phi_c =1.325 \pm 0.05$, respectively.\\

According to the numerical studies by Petchenko et al. \cite{petchenko2006violent,petchenko2007flame}, the large-amplitude oscillations and flame folding would disappear for $h/\delta_T<25$. In our experiments, they were not observed  for $h/\delta_T<200$. The small-amplitude oscillations, characteristic of the primary acoustic instability, also ceased for $80< h/\delta_T < 140$, a behavior not reported in \cite{petchenko2007flame} but in agreement with the predictions by Kurdyumov and Matalon \cite{Kurdyumov2016} in the limit $h/\delta_T \ll 1$.  \\
\noindent

Viscous effects may also account for Saffman-Taylor instabilities in narrow-channel flows. Subtle differences on the acoustic stability were found in the experimental study by Aldredge in \cite{aldredge2004saffman} and \cite{aldredge2005methane} with methane flames and in the numerical work by Kang et al. \cite{kang2003computational} for sufficiently low P\`eclet numbers. The variation of viscosity across the flame implies additional forces that interact with the flame front, whose thickness and characteristic velocities are slightly modified. These effects are hardly discerned in our experiments, probably disguised in the whole set of instabilities described here for the widest channels tested, as can be seen in Fig. \ref{fig:DME_lateral}. Nevertheless, even after suppressing the acoustic oscillations by reducing the gap of the cell to a minimum of $h=1$ mm, when we reached a nearly-viscous DME flow, the typical Saffman-Taylor structures were not observed either.

\section{Discussion}
The experimental observations of the flame interacting with the acoustic pressure waves traveling in a tube encouraged several authors to investigate the physical reasons behind that behavior. Yoon et al. \cite{yoon2016onset} reported a correlation between the average acoustic intensity and the coupling constant $\beta M$. According to their results, the primary instability would only develop if $\beta M$ is larger than a critical value. In all the mixtures tested in our experiments with $h=10$ mm, we observed the vibration of the flame and, assuming the conclusion in \cite{yoon2016onset} as correct, we can infer that the coupling constant $\beta M$ included in Table~\ref{table:flame_parameters} is above its critical value when the acoustic losses were negligible. As we showed above in Fig.~\ref{fig:DME_position_h}, the acoustically-induced flame oscillations disappeared in narrow channels $h<7 $ mm as a consequence of the viscous attenuation of the acoustic pressure waves, a result that Yoon et al. \cite{yoon2016onset} also identified in their work.\\ 

According to our experiments, the transition to the secondary instability takes place only for $\phi > \phi_c= 0.87 \pm 0.05$ in propane flames, for $\phi < \phi_c=0.95 \pm 0.05$  in methane flames and for $\phi> \phi_c=0.55 \pm 0.05$ in DME flames. The reverse influence of the equivalence ratio on the transition from the primary to the secondary instability observed in propane, DME and methane flames cannot be explained in terms of flame temperature $T_b$ or burning velocity $S_L$. As can be easily checked in Table \ref{table:flame_parameters}, these two parameters have a non-monotonic dependence on $\phi$, with similar values of $T_b$ and $S_L$ in rich and lean flames that do not anticipate the differences observed in the experiments.\\

Encouraged by the work of Aldredge and Killingsworth \cite{aldredge2004experimental} with methane-air flames, we calculated in Table \ref{table:flame_parameters} the corresponding values of the effective Lewis number $Le_{eff}$, defined as in \cite{matalon2009flame}, and the values of the Markstein number $\mathcal{M}$ calculated numerically by Bechtold and Matalon \cite{bechtold2001dependence} for methane and propane, and calculated in this work for DME from the measurements of the Markstein length carried out by de Vries et al.~\cite{de2011laminar}. Using these data, we see that the onset of the secondary instability is only observed in the experiments for $\cal{M}$ below a critical value:  ${\cal M}_c\simeq 2.92 $ for propane flames, ${\cal M}_c \simeq 0.82 $ for methane flames and ${\cal M}_c \simeq 3.90$ for DME flames. This fact, together with the decrease of the Markstein number with $\phi$ towards rich (propane and DME) or lean (methane) mixtures, appoints it as the possible controlling parameter of the empirical observations.\\

Matalon and Metzener \cite{matalon1997propagation} performed a theoretical and numerical study of premixed flames propagating in closed tubes. Particularly relevant for the experimental study presented here is the stability analysis in terms of the Markstein number, the only mixture-sensitive parameter of their model. According to their results, the shape of the flame corrugates for values of ${\cal M < \cal M}_c$, amplifying the perturbations in a range of wavelengths determined by  ${\cal M}$. The flame then assumes either a cellular shape, characterized by a finite number of cells convex toward the unburned gas, or a tulip shape. The critical Markstein number ${\cal M}_c$ depends on the aspect ratio $L/h$ of the channel. If the initial deformation of the flame front is a consequence of the hydrodynamic and thermo-diffusive instabilities, the form adopted later on during the propagation is due to nonlinearities, hard to describe using analytical methods. The Markstein number will determine which perturbations' wavelengths are amplified \cite{matalon1997propagation} and, therefore, the final shape of the reactive surface.\\ 

During its propagation, the flame is an important source of acoustic noise. The interaction of the noise with the corrugated surface of the flame triggers a coupling mechanism that leads to the high-amplitude oscillations described above for methane, propane and DME. The reverse behavior of propane and DME flames with respect to that of methane and, apparently, hydrogen \cite{yang2015oscillating}, matches with the variation of the Markstein number with equivalence ratio. Also, the presence of a critical Markstein number, given above for the different fuels, seems to confirm the importance of the flame shape in the development of the secondary thermo-acoustic instability. 

\section{Conclusions}
Laminar premixed methane-air, propane-air and DME-air flames propagating inside a Hele-Shaw cell were studied experimentally in order to examine the transition from the primary to the secondary acoustic instability. The primary acoustic instability is characterized by small-amplitude flame oscillations, which flatten but do not change significantly the shape of the reaction front. During the whole propagation, as it is shown for first time in Fig. \ref{fig:lateral}, the shape of the flame in the transversal z-direction remains parabolic and convex towards the fresh mixtures.\\
The secondary acoustic instability is defined by large-amplitude flame oscillations. For the equivalence ratios tested in our experiments, they appear in flames richer than a critical value for propane-air ($\phi_c=0.87$) and DME-air ($\phi_c=0.55$) mixtures and in flames leaner than $\phi_c=0.95$ for methane-air mixtures. When at this regime, the outline of the flame-front changes drastically, forming pulsating finger-shaped structures and reaching high oscillatory accelerations due to the acoustic-flame coupling. In all the fuels tested, the onset of the pulsations is observed for Markstein numbers $\mathcal{M}$ below a critical value $\mathcal{M}_c$, confirming the numerical results by Matalon and Metzener \cite{matalon1997propagation}. Besides, its decrease towards rich (propane and DME) and lean (methane) flames makes it the best candidate to characterize the transition observed in the experiments.\\

The lateral view recordings depict the three-dimensional structure of the flame. During the propagation in the primary acoustic instability, the shape of the flame is parabolic and remains that way during the small-amplitude oscillations of the primary instability. On the other hand, when $\mathcal{M}<\mathcal{M}_c$, just before the transition to the secondary instability, the shape of the flame transverse to the direction of flame propagation abandons the parabolic profile that adopted after the ignition to become much flatter. Immediately after that, the reaction region forms small-size bulges that grow to form the finger-like cell structure that characterizes the secondary instability.\\

The effect of the channel thickness $h$ was also assessed in this paper for stoichiometric DME flames. The reduction of $h$ diminished the thermo-acoustic instability as a consequence of the viscous damping of the acoustic wave traveling along the channel. The transition from the secondary to the primary oscillatory regime for a stoichiometric flame occurred between $10$ mm $< h < 7$ mm. Farther decrease on $h$ completely eliminates any oscillatory instability at the reaction zone. Much richer flames are needed to capture the transition in configurations with $h=7$ mm ($\phi_c=1.175 \pm 0.05$) and $h=4$ mm ($\phi_c=1.325 \pm 0.05$) than for $h=10$ mm ($\phi_c=0.55 \pm 0.05$). The transition to the secondary instability was not observed in the narrowest channel ($h=1$ mm).

 \section*{Acknowledgements}
\noindent
This work was supported by projects ENE2015-65852-C2-1-R
(MINECO/FEDER, UE) and BYNV-ua37crdy (Fundaci\'on Iberdrola Espa\~na). The authors wish to thank the technical knowledge and  assistance of Y. Babou, D. D\'iaz, I. Pina,  and M. Santos in the design, construction and operation of the experimental setup.
\let\itshape\upshape
\bibliographystyle{elsarticle-num}
\bibliography{paperbib_tuneao}
\addcontentsline{toc}{section}{References}

\newpage

\end{document}